\documentclass[%
 reprint,
 amsmath,amssymb,
 aps,
]{revtex4-2}

\usepackage{graphicx}
\usepackage{dcolumn}
\usepackage{bm}
\usepackage{hyperref}

\usepackage{xcolor}
\begin{document}

\preprint{APS/123-QED}

\title{Social patch foraging theory in an egalitarian group}

\author{Lisa Blum Moyse}
\author{Ahmed El Hady}%
 \altaffiliation[Also at ]{Department of Collective Behaviour, Max Planck Institute of Animal Behaviour, Konstanz, Germany} 
\affiliation{%
Centre for the Advanced Study of Collective behavior,  University of Konstanz, Konstanz, Germany 
}%




\date{\today}

\begin{abstract}
Foraging is a widespread behavior, and being part of a group may bring several benefits compared to solitary foraging, such as collective pooling of information and reducing environmental uncertainty. Often theoretical models of collective behavior use coarse-grained representations, or are too complex for analytical treatment, and generally do not take into account the noisy decision making process implemented by individual agents. This calls for the development of a mechanistic, analytically tractable, and stochastic framework to study the underlying processes of social foraging, tying the microscopic to the macroscopic levels. Based on an evidence accumulation framework, we developed a model of patch-leaving decisions in a large egalitarian group. Across a variety of environmental statistics and information sharing mechanisms, we were able to analytically derive optimal agent strategies. The environmental statistics considered are either two non-depleting or several successive depleting patches. The social information sharing mechanisms are either through observation of others' food rewards or through belief sharing, with continuous sharing, pulsatile observation of others' departures or arrivals, or through counting the number of individuals in a patch. Throughout all these conditions, we quantified how cohesive a group is over time, how much time agents spend on average in a patch and what are their group equilibrium dynamics. We found that social coupling strongly modulates these features across a variety of environmental statistics.  This general modeling framework is crucial to both designing social foraging experiments and generating hypotheses that can be tested. Moreover, this framework can be extended to groups exhibiting hierarchical relations.  

\end{abstract}

\maketitle

\makeatletter
\def\l@paragraph{\@dottedtocline{5}{5.3em}{2.1em}}
\makeatother

\tableofcontents

\section{Introduction}
Foraging is a widespread decision making behavior~\cite{stephensForagingBehaviorEcology2007}. Animals often forage socially, like eusocial insects~\cite{detrainCollectiveDecisionMakingForaging2008}, birds~\cite{aplinIndividuallevelPersonalityInfluences2014}, rats~\cite{alfaroRoleOutcomeUnit2019}, meerkats~\cite{gallGroupCohesionForaging2017}, or baboons~\cite{strandburg-peshkinSharedDecisionmakingDrives2015}.
While social foraging affects resource availability~\cite{fernandez-juricicFlockDensitySocial2004}, being part of a group brings several benefits compared to solitary foraging, such as safety against predators~\cite{siegfriedFlockingAntipredatorStrategy1975}, enhanced ability to capture prey~\cite{dumkeAdvantagesSocialForaging2018}, or collective pooling of information~\cite{clarkEvolutionaryAdvantagesGroup1986}\cite{pitcherFishLargerShoals1982}. Indeed for the latter, social information sharing may lead to an increased foraging success~\cite{martinez-garciaOptimizingSearchResources2013}\cite{dallInformationItsUse2005}.\\
Mechanisms underlying social interactions can be divided into two categories: environmental information actively shared via specific signals between agents, or social information observed by others~\cite{leadbeaterSocialLearningInsects2007}\cite{galefSocialInfluencesForaging2001}. An example of active signals are vocalizations, used to communicate in species such as capuchin monkeys~\cite{boinskiVocalCoordinationTroop1993} or bats~\cite{kohlesSociallyForagingBats2020}. In bees, information about patch locations can be communicated through the waggle dance~\cite{frischDanceLanguageOrientation2013}. Another mechanism, trophallaxis-- the exchange of liquid food by mouth-- allows honeybees to learn associations between nectar odors and rewards~\cite{gilOlfactoryLearningMeans2005}.In addition to these directed signals, social foraging relies also on information-bearing cues~\cite{galefSocialInfluencesForaging2001}. Inside this category, two social information types can be distinguished: social cues, which exposes the internal belief of an individual, and public information, when the behavior of another agent reveals information about the environment.
For example, the presence of other individuals constitutes a social cue that may attract the observer to a patch location. This process is known as local enhancement~\cite{leadbeaterSocialLearningInsects2007}. Examples of this phenomenon can be found in various species, such as birds that tend to move to places where other birds are feeding~\cite{waiteCopyingForagingLocations1988}, or in Norway rats, that follow each other from their burrows to food or water locations~\cite{calhounEcologySociologyNorway1963}.
Another social cue example is chemical trails left by foraging bees, that can lead others' to patches~\cite{niehRecruitmentCommunicationStingless2004}.The other type of inadvertent cue, public information, has for example been observed in  starlings~\cite{templetonPatchAssessmentForaging1995}, where the observation of others' foraging success gives information on the current patch~\cite{valonePatchEstimationGroup1993}.
\\

To investigate the underlying processes of social foraging, several theories and quantitative models have been developed. A theory like the ideal free distribution predicts the distribution of foraging individuals among patches with different food amounts so that fitness is optimized and competition is minimized~\cite{fretwellTerritorialBehaviorOther1969}. However empirical data often show a deviation from these predictions~\cite{kennedyCanEcologicalTheory1993}. Another model uses the  marginal value theorem, modified to study groups' residence times, compared to solitary ones~\cite{livoreilPatchDepartureDecisions1997}.
Agent-based models are a useful framework to implement complex individual and interaction rules~\cite{wajnbergOptimalPatchMovement2013}, as well as reinforcement learning models~\cite{falcon-cortesCollectiveLearningIndividual2019}\cite{lofflerCollectiveForagingActive2023}. Some game-theoretic approaches have also been applied to social  foraging~\cite{giraldeauSocialForagingTheory2000} \cite{cressmanGameTheoreticMethodsFunctional2014}, which allows studying 
the competitive and cooperative behaviors between agents, or the impact of a group's size. The way agents integrate both their own and social information to increase their knowledge about their environment has been studied in Bayesian models~\cite{perez-escuderoCollectiveAnimalBehavior2011}. Movement decisions in groups can also be studied theoretically, to understand the impact of the proportion of informed individuals~\cite{couzinEffectiveLeadershipDecisionmaking2005}.\\

The aforementioned approaches however are often coarse-grained, implying general interaction rules, or are too complex to be treated analytically. Another limitation is that uncertainty and noise in decision processes are often not taken into account, which can strongly shape the agents' patch departure statistics~\cite{davidsonForagingEvidenceAccumulation2019}\cite{kilpatrickUncertaintyDrivesDeviations2021a}. This calls for the development of an analytically tractable framework, including inherent stochasticity, to study mechanisms underlying social foraging across a variety of social and environmental conditions. Models developed within this framework should also be amenable to be fitted to experimental data to unravel mechanisms underlying the decision making of animals while they forage socially. This paper introduces a new framework to describe quantitatively collective patch foraging dynamics, and will focus here on collaborative foraging.\\

This new model is built on a widespread stochastic decision-making framework based on the evidence accumulation process, 
drift-diffusion models, which can be applied to patch-leaving tasks~\cite{davidsonForagingEvidenceAccumulation2019}. Such processes have related neural correlates~\cite{liuNeuralCorrelatesEvidence2011}\cite{hukNeuralCorrelatesNeural2012}, and have been useful 
to understand the neural basis involved in binary perceptual choices~\cite{goldNeuralBasisDecision2007}.
In these models, each individual accumulates information about a patch quality, integrated into a decision variable that represents the current agent's belief. When this variable reaches a fixed threshold, the decision to leave the patch is taken~\cite{davidsonForagingEvidenceAccumulation2019}. 
Decision times can then be calculated as the solutions to first passage time problems using stochastic processes and asymptotics~\cite{gardinerHandbookStochasticMethods1985}\cite{bogaczPhysicsOptimalDecision2006}. 
Such mechanistic models have for now only focused on one~\cite{davidsonForagingEvidenceAccumulation2019} or two foragers~\cite{bidariStochasticDynamicsSocial2022}. The last model introduced a theoretical framework to study coupled agents for different interaction mechanisms~\cite{bidariStochasticDynamicsSocial2022}. In this paper, this model is generalized to larger groups, with more interaction mechanisms, across environments with different statistics.\\

The two interaction mechanisms described in~\cite{bidariStochasticDynamicsSocial2022} are the diffusive and pulsatile coupling. The first one constitutes a continuous sharing of information, studied in some models of ``ideal groups'' of decision-makers~\cite{srivastavaCollectiveDecisionMakingIdeal2014}\cite{sorkinSignaldetectionAnalysisGroup2001}, as well as in models of migrating animal groups~\cite{paisAdaptiveNetworkDynamics2014}\cite{torneySpecializationEvolutionaryBranching2010}. For the second one, pulsatile coupling, information is shared only when a departure decision is made~\cite{caginalpDecisionDynamicsGroups2017}\cite{karamchedBayesianEvidenceAccumulation2020}. Diffusive and pulsatile coupling can be considered respectively as an actively shared signal and as a social cue~\cite{dallInformationItsUse2005}\cite{leadbeaterSocialLearningInsects2007}\cite{galefSocialInfluencesForaging2001}.
In this paper, new interaction mechanisms are introduced. First, pulsatile coupling is extended to account also for pulses of information when another agent enters the patch. A second new mechanism is counting coupling, which represents another way to notice the information associated with others' presence. Here the number or proportion of other individuals in the same patch represents the socially relevant information. The two processes may also  be characterized as social cues. A last introduced mechanism, reward coupling, that represents the observation of others' food catches, constituting a form of  
public information.\\

These social interaction mechanisms are studied within environments of different statistics, that allow studying different collective decision-making aspects. Three patch organizations are studied: a single patch, two non-depleting patches, and multiple successive depleting patches. The first environment is useful to lay the basis of the different tools and metrics that are used in the subsequent cases. The metrics used to characterize a group's behavior are the probability of leaving a patch, and the fraction of agents in a patch. From these metrics, three main features are extracted, related to key social aspects: cohesion, accuracy, and exploitation~\cite{franksSpeedCohesionTradeoffs2013}\cite{stroeymeytImprovingDecisionSpeed2010}. Throughout this paper, cohesion will refer to individuals remaining together in a same patch, accuracy to the proportion of agents in the highest quality patch, and exploitation to the time spent foraging in a patch.\\
A condition with all agents initially in the same patch enables the quantification of the group's cohesion. This can be observed with the damping of oscillations of the probability distributions. The second aspect, accuracy, is here defined as the fraction of agents in the best patch. This can be evaluated in the two non-depleting patches environment.
Finally, in an environment with successive depleting patches, the time spent in a patch characterizes to which degree that patch is exploited. The mean residence time is a useful metric to quantify this aspect.
Throughout this paper, numerical simulations and analytical analysis enable characterizing how these different information sharing mechanisms and environments lead to various collective dynamics.\\

\section{Methods}
\subsection{Model}
Agents move between different patches identified by a number $k$ in case of a several patches environment. For a two-patches environment in section~\ref{twopatchessec}, $k=0$ is the initial patch, and $k=1$ is the highest quality patch. For successive patches in section~\ref{Succpatches}, the patches are numbered from $1$ to $K$, the total number of patches. The decision to leave a patch for another one is taken through the underlying process of evidence accumulation.\\

$x_i$, the decision variable of a foraging agent $i$ in a patch $k_i$, evolves according to the following stochastic differential equation

\begin{equation}
dx_i(t)=(r_i(t)-\alpha)dt + \sum_{j\neq i} c_{ij}(t)dt+\sqrt{2B} dW_i(t)
\label{x}
\end{equation}

with the initial condition $x_i(0)=0$. The forager leaves a patch when $x_i$ reaches the threshold $\theta$. The travel time between patches is $T_\text{tr}$. The drift-diffusion process is illustrated in Fig.~\ref{fig:Panel1}(a) or Fig.~\ref{fig:Panel2}(a), for a single or two patches environment.\\
The decision variable evolves following an evidence accumulation process, where $\alpha$ represents the cost associated with foraging, and $r_i(t)$ the food rewards.
Every time step $\Delta_{r}$, an agent has a probability $p^{k_i}$ of getting a food reward $r_i=1$. This probability may be constant $p^{k_i}=p^{k_i}_\text{in}$ (non-depleting environment), or decrease as the amount of food in a patch is consumed (depleting case). In the latter case, $p^{k_i}(t+dt) = p^{k_i}(t) - \frac{\sum_i r_i(t)}{A_m^{k_i}}$, with $A_m^{k_i}$ the maximum amount of food in a patch $k_i$.\\
$W(t)$ is the standard Wiener process, $B$ represents the noise amplitude, and the term $c_{ij}$ corresponds to the interaction mechanisms, described below.\\

Eq.~\eqref{x} and the reset of $x_i$s at every departure do not include a memory effect throughout the different patches. In the two patches section, see~\ref{IndLearning}, agents with an internal long-term belief of which patch has the highest quality are studied.\\

\textit{Best patch inference}\\
To represent this long-term belief dynamics, the variable $y_i$ is introduced. 
This variable represents the inference of which patch has the highest reward probability $p^k$. It evolves as
\begin{equation}
\tau_y dy_i = (k_i-y_i)r_i(t)dt
\label{y}
\end{equation}
These dynamics depend here only on the individual food catches. At $t=0$, $y_i(0) = \frac{1}{2}$. $y_i$ converges then to $k=0$ or $k=1$. The more rewards are received in a patch $k$, the faster the convergence is.
$\tau_y$ is the timescale of the differential equation.\\

When an agent $i$ is leaving a patch at time $t$, the factor $\omega^{k_i}$ is updated with  $y_i(T)$:
\begin{equation}
\omega^{k_i} = \frac{y_b+y_i(t)(1-2k_i) + k_i}{y_b+\frac{1}{2}} 
\end{equation}
modifying the $\alpha$ parameter as $\alpha \xrightarrow{} \alpha\times \omega^{k_i}$.
For example, if $y_i(t)>\frac{1}{2}$ (i.e. the estimated best patch is $k=1$), $\omega^{k_i}<1$ in patch $k=1$ and $\omega^{k_i}>1$ in patch $k=0$. So $\alpha\times \omega^{k_i}$ is reduced in $k=1$, implying that the agent stays longer in this estimated best patch. And inversely for the other patch $k=0$.\\

This long-term memory effect is only studied for non-interacting agents in this paper. Interaction rules for social agents are described below.\\

\textit{Social coupling}\\
Information sharing from individual $j$ to $i$ is expressed by the coupling term $c_{ij}$, decomposed in two different terms such as 
\begin{equation}
    c_{ij}(t)=v_{ij}(t)\times\bigl(a_{ij}(t)+b_{ij}(t)\bigr)
\end{equation}
$v_{ij}(t)$ is a vector equal to $1$ if agents $i$ and $j$ are in the same patch, equal to $0$ otherwise.\\
The different coupling types are detailed below and are illustrated in Fig.~\ref{fig:Panel3}A.\\ 

Information can first come from the observation of food rewards that other agents in the same patch get.\\ If the number of agents in a patch $k_i$ ($n^{k_i}(t)$) respects $n^{k_i}(t)>1$,
\begin{equation}
    a_{ij}(t) = \kappa_r\frac{r_j(t)}{n^{k_i}(t)-1} 
\end{equation}
$\kappa_r$ corresponds to reward coupling strength.\\

Individuals can also share their beliefs, we distinguish three different types of coupling.\\

\begin{itemize}
    \item First, agents can share their beliefs continuously with other agents in the same patch. This is diffusive coupling.
\begin{equation}
    b_{ij}(t) = \frac{\kappa_\text{diff}}{N} \biggl(x_j(t)-x_i(t)\biggr)
\end{equation}

The parameter $\kappa_\text{diff}$ is scaled by the size of the group $N$. While this parameter can also be scaled by the number of agents in a current patch $n^k(t)$, this case is not analytically tractable. Numerical simulations show similar results for the last case, see supplementary Fig~\ref{fig:Supdiff}.

\item Agents can also perceive how many individuals in their group are in the same patch. This is counting coupling. Without the sum in eq.~\eqref{x},
\begin{equation}
    b_{i}(t) = \kappa_c \biggl(\frac{n^{k_i}(t)}{N}-\eta\biggr)
\end{equation}
If many agents are in the same patch, $n^{k_i}(t)>\eta$, so $b_{i}>0$, i.e the agent accumulates evidence to stay in the patch. Inversely for $n^{k_i}(t)<\eta$.

\item Agents can also communicate through a pulse of information, corresponding to an agent's arrival ($\kappa_a$) or departure ($\kappa_d$) in a patch. This is pulsatile coupling.\\
If two agents were in the same patch at time $t$, and the agent $j$ is leaving:
\begin{equation}
b_{ij}(t)= -\frac{\kappa_d}{\mathcal{N}_d}\times\frac{\delta(x_j(t)-\theta)}{dt}
\label{eqkd}
\end{equation}

If the agent $j$ is in the previous patch, leaving to join the current patch of agent $i$:
\begin{equation}
b_{ij}(t)= \frac{\kappa_a}{\mathcal{N}_a}\times\frac{\delta(x_j(t-T_\text{tr})-\theta)}{dt} 
\label{eqka}
\end{equation}

The normalization terms are equal either to $\mathcal{N}_d=\mathcal{N}_a=N$, or $\mathcal{N}_d = n^{k_i}-1$, $\mathcal{N}_a = N-n^{k_i}-1$.

\end{itemize}

\begin{table}[h!]
\begin{tabular}{|c|c|} 
 \hline
 Variable or parameter & Definition \\ [0.5ex] 
 \hline
 $i$ & Index of an agent\\
 $k$ & Index of a patch\\
 $x_i$ & Decision variable  \\ 
 $\alpha$ & Cost of foraging  \\ 
 $r_i(t)$ & Food reward \\ 
 $B$ & Noise parameter \\ 
 $W(t)$ & Wiener process \\ 
 $\theta$ & Decision threshold \\ 
 $T_\text{tr}$ & Travel time between patches \\ 
 $p^k(t)$ & Probability of reward \\
 $p_\text{in}^k(t)$ & Initial probability of reward \\
 $A_m^k$ & Maximum amount of food in a patch \\
 $\Delta_r$ & Time step of rewards \\ 
 $n^k(t)$ & Number of agents in a patch \\
 $c_{ij}(t)$ & Global coupling term \\
 $v_{ij}(t)$ & Same patch term\\
 $a_{ij}(t)$ & Reward coupling term \\
 $\kappa_r$ & Coupling reward parameter \\
 $b_{ij}(t)$ & Belief coupling term \\
 $\kappa_\text{diff}$ & Coupling diffusive parameter\\
 $\kappa_d$ & Coupling departure parameter \\
 $\kappa_a$ & Coupling arrival parameter \\
 $\kappa_c$ & Coupling counting parameter \\
 $\eta$ & Threshold counting parameter \\
 $y_i$ & Best patch inference\\
 $\tau_y$ & Best patch inference timescale\\
 $y_b$ & Boundary  $y$ factor\\
 $\omega^{k_i}$ & Best patch inference factor\\
 \hline
\end{tabular}
\caption{Variable and parameter definitions of the model}
\label{table:1}
\end{table}

\begin{table}[h!]
\begin{tabular}{|c|c|} 
 \hline
 Variable or parameter & Definition \\ [0.5ex] 
  \hline
 $T$ & Patch residence time \\
  $\overline{T}$ & Mean $T$ (exploitation)\\
  $\widetilde{\alpha}^k(t)$ & Effective drift term\\
 $\Psi^k(T)$ & Probability distribution of $T$s \\
 $\Omega^k(T)$ & Probability to be in the patch at $T$\\
 $P^k(t)$ & Probability to leave the patch at $t$ \\
 $Q^k(t)$ & Probability to be in the patch at $t$\\
 $\widetilde{\alpha}_\text{eq}^k$ & $\widetilde{\alpha}^k(t)$ at equilibrium\\
 $P_\text{eq}^k$ & $P^k(t)$ at equilibrium\\
 $Q_\text{eq}^k$ & $Q^k(t)$ at equilibrium (accuracy)\\
 $\gamma$ & Damping parameter (cohesion)\\
 $\overline{y}(t)$ & Estimated best patch inference\\
 \hline
\end{tabular}
\caption{Variable and parameter definitions of the analysis}
\label{table:2}
\end{table}

\begin{table}[h!]
\begin{tabular}{|c|c|} 
 \hline
 Changing element & Type \\ [0.5ex] 
  \hline
 Environment &  1/ 2 / K successive patches \\
  Variation of food amount & Non-depleting/Depleting\\
  others' rewards information & Without/ With\\
  Belief sharing & No coupling/ Diffusive/ \\
  &  Pulsatile (1,2)/ Counting\\
 \hline
\end{tabular}
\caption{Table of all conditions}
\label{table:3}
\end{table}

\subsection{Simulations details}
The stochastic differential equation~\eqref{x} of the decision variable is solved numerically with an Euler method, with $dt=0.01$ s. Other variables are updated at every time step.\\

\textit{Parameters:}
\begin{itemize}
    \item In Fig.~\ref{fig:Panel1}, parameters are $\theta=-5$, $\alpha=1.25$, $B=0.1$, $p_\text{in}=0.8$, $A_m=100$, $\Delta_r=0.02$ s.
    
    \item In figures of the two patches section~\ref{twopatchessec}, except in theoretical plots,  non-social parameters remain constant: $\theta=-5$, $\alpha=1.25$, $B=0.1$, $\Delta_r=0.01$ s, $p^0=0.4$, $p^1=0.6$. $T_\text{tr}=1$s in Fig.~\ref{fig:Panel2} and $T_\text{tr}=0$s otherwise. For damped oscillations $\gamma$ plots, $p^0=p^1=0.5$.

    \item In Fig.~\ref{fig:Panel5}, numerical parameters are $\theta=-5$, $\alpha=1.25$, $B=0.1$, $\Delta_r=0.01$ s, $K=10$, $N=10$, $A^k_m = 8000$, $p^k_\text{in}=\frac{2}{3}$ for all $k$.
    
\end{itemize}

In figures, blue histograms from the numerical simulations corresponding to distributions of leaving times $P^k(t)$ are normalized according to the equilibrium probabilities $P^k_\text{eq}$.\\ 
\section{Results}
Patch-leaving dynamics in the three different environments studied in this paper are presented in the following sections: a single patch, two-non depleting patches, and successive depleting patches.

\subsection{Single patch, single agent}
\begin{figure*}
    \centering
    \includegraphics[width=17cm]{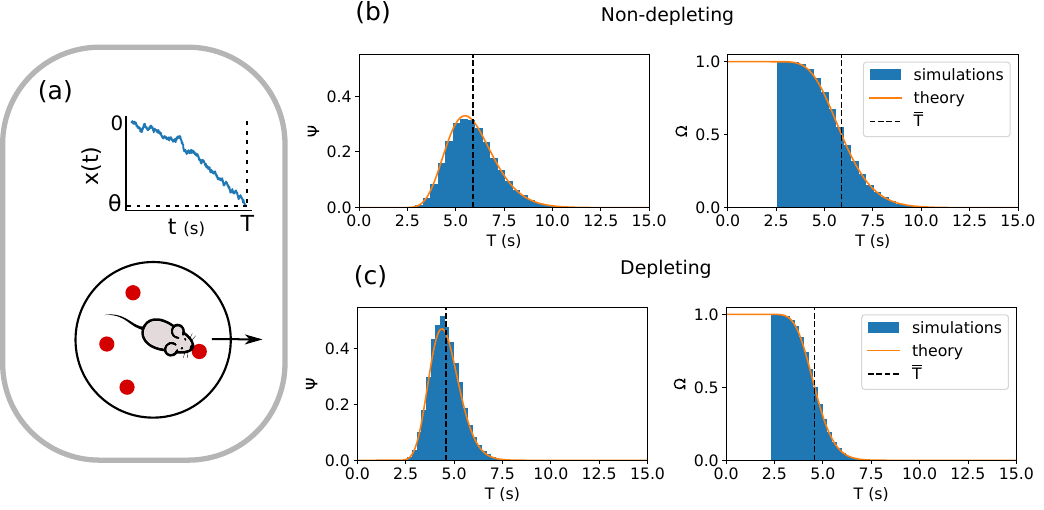}
    \caption{{\textbf{Single agent foraging in a single patch}}. (a) Single patch foraging task: A foraging agent accumulates evidence of the rate of food rewards in the patch. The higher this rate is, the slower the decision variable $x(t)$ decreases, and the longer this agent tends to stay in the patch. The decision to leave is made when the decision threshold $\theta$ is reached by the decision variable. (b), (c) Probabilities of leaving the patch $\Psi$ (Left), and being in the patch $\Omega$ (Right), for a non-depleting (b) and depleting (c) patch. The blue histograms are the distributions from the numerical simulations ($5\times 10^4$ simulations). The curves correspond to the theoretical expressions~\eqref{Psi},~\eqref{Omega}, with average rates $\overline{p}$ given by equation~\ref{p} (non-depleting) or~\eqref{pdep} (depleting). The mean departure time $\overline{T}$ is reduced in the depleting case. }
    \label{fig:Panel1}
\end{figure*}

In this section the simple situation of a single individual foraging in a patch is studied, schematized in Fig.~\ref{fig:Panel1}(a). This part lays the basis that is used for more complex situations, with more patches, interaction mechanisms between agents, and long-term memory effects. The index $k$ is removed in this section for readability reasons.\\
Throughout the article, the reward time step $\Delta_r$ is taken as small ($\Delta_r<<1$). Thus, in this quasi-continuous reward approximation, it is possible to replace the reward term $r(t)$ by their average rate $\overline{p}(T)$ and to consider the effective drift term $\widetilde{\alpha}(T)=\alpha-\overline{p}(T)$. This average rate $\overline{p}(T)$ has different expressions, depending on the non-depleting or depleting feature of a patch.\\

\paragraph{Non-depleting environment:}
In this case, the average rate can be written as
\begin{equation}
    \overline{p}=\frac{p dt}{\Delta_r} \label{p}
\end{equation}
If $\widetilde{\alpha}>0$, the mean patch residence time $\overline{T}$ can be obtained by integrating equation~\eqref{x} from $0$ to $\theta$.

\begin{equation}
\overline{T}=-\frac{\theta}{\widetilde{\alpha}}
\label{T}
\end{equation}

With this constant drift $\widetilde{\alpha}$, the probability of leaving a patch at time $T$ is an inverse Gaussian distribution~\cite{coxTheoryStochasticProcesses2017}

\begin{equation}
\Psi(T) = -\frac{\theta}{\sqrt{4\pi B T^3}}\exp \biggl(\frac{-(\theta+\widetilde{\alpha} T)^2}{4BT} \biggr)
\label{Psi}
\end{equation}

And the probability to be in the patch at a time $T$ is

\begin{equation}
\Omega(T) =  \Phi\biggl(\frac{-\theta-\widetilde{\alpha} T}{\sqrt{2BT}}\biggr) - \exp\biggl(\frac{-\widetilde{\alpha}\theta}{B}\biggr)\Phi\biggl(\frac{\theta-\widetilde{\alpha} T}{\sqrt{2BT}}\biggr)
\label{Omega}
\end{equation}

With the standard normal integral:
\begin{equation}
    \Phi(y)= \frac{1}{\sqrt{2\pi}}\int_{-\infty}^y \exp\biggl(-\frac{z^2}{2}\biggr)dz
\end{equation}

Contrary to a non-depleting patch, in a depleting environment the average rate is not constant.

\paragraph{Depleting environment:}
We can write the evolution of the probability to get a reward as
 $p(t)=\frac{p_\text{in} dt}{\Delta_r} \times e^{-\frac{t}{A_m \Delta_r}}$. The average reward rate is estimated as
\begin{equation}
\overline{p}(T) = \frac{\int_0^{T} \frac{p_\text{in} dt}{\Delta_r} e^{-\frac{t}{A_m \Delta_r}}dt}{T}=p_\text{in}dt A_m \frac{1-e^{-\frac{T}{A_m\Delta_r}}}{T}
\label{pdep}
\end{equation}
This term $\overline{p}(T)$ can be used in equations~\eqref{T} to~\eqref{Omega}.

To find the value of $\overline{T}$, the self-consistent equation~\eqref{T} should be solved numerically.\\

These two probability distributions $\Psi(T)$ and $\Omega(T)$, as well as the mean patch residence time $\overline{T}$ are shown in Fig.~\ref{fig:Panel1}(b) and \ref{fig:Panel1}(c) (respectively for the non-depleting and depleting case). The distributions are shifted to earlier departures in the depleting case, with a reduced mean departure time $\overline{T}$.\\

This section established the analytical basis which will be further used and developed in the next section, studying different collective foraging dynamics in a two non-depleting patches environment.

\subsection{Two non-depleting patches}
\label{twopatchessec}
In this section, one or several foraging agents can move between two non-depleting patches of food ($k=0$ and $k=1$), with different reward probabilities $p^0$ and $p^1$. The initial patch for all agents is the patch $k=0$. Fig.~\ref{fig:Panel2}(a) represents a foraging agent in this two patches environment and the corresponding evidence accumulation process. A travel time $T_\text{tr}$ separates the two patches. Every time an agent decides to leave for the other patch, their decision variable $x_i$ is reset to zero. Two different social conditions are studied successively in this section: Single or non-interacting agents (without~\ref{2single} or with best patch inference~\ref{IndLearning}), and a group of interacting agents~\ref{socinfo}.

In particular, the first section below enables to define the different distributions and metrics (cohesion, accuracy, exploitation) to characterize the collective dynamics.

\subsubsection{Characterization of distributions for non-interacting agents}
\label{2single}
The probability to leave a patch $k$ at time $t$, $P^k(t)$, is given by the convolution over the $\Psi^0(t)$ and $\Psi^{1}(t)$. This probability can be found after using the Laplace transform of $\Psi$ to compute the convolution of functions with the same drift term in one compartment (see Appendix~\ref{App1}).

\begin{equation}
P^k(t) = \sum_{\nu=1}^\infty \bigl(\Psi^0_{\nu}*\Psi^1_{\nu-1+k}\bigr)\bigl(t-(2\nu-2+k) T_\text{tr}\bigr)
\label{Pk}
\end{equation}
For $t\leq (2\nu-2+k)T_\text{tr}$, the convolution is equal to zero.\\

With
\begin{equation}
    \Psi^k_{\nu}(t)=\frac{-\nu\theta}{\sqrt{4\pi B t^3}} \exp \biggl(\frac{-(\nu\theta+\widetilde{\alpha}^k t)^2}{4Bt} \biggr)
    \label{Psinu}
\end{equation}
for $\nu\geq 1$. $\Psi^k_0(t) = 1$ for all $t$.\\
$\widetilde{\alpha}^k$ is the effective drift term, defined here as $\widetilde{\alpha}^k = \alpha-p^k$.

The probability to be in the patch at a time $t$ is given by the time integral of the arrival minus the departure probability of a patch $k$:
\begin{equation}
    Q^k(t) = 1-k + \int_0^t d\tau [P^{k'}(\tau-T_\text{tr})-P^k(\tau)] 
    \label{Qk}
\end{equation}

In the special case $p^1=p^{0}$, the distributions are
\begin{equation}
   P^k(t) = \sum_{\nu=1}^\infty \Psi^k_{2\nu-1+k}\bigl(t-(2\nu-2+k) T_\text{tr}\bigr)
   \label{Pequal}
\end{equation}
and
\begin{equation}
\begin{split}
   Q^k(t) = \sum_{\nu=1}^\infty& \biggl[\Omega^k_{2\nu-1+k}\bigl(t-(2\nu-2+k) T_\text{tr}\bigr)\\&-\Omega^{k'}_{2\nu-1+{k'}}\bigl(t-(2\nu-1+{k'}) T_\text{tr}\bigr)\biggr]
\end{split}
\end{equation}

With
\begin{equation}
\Omega^k_\nu(T) =  \Phi\biggl(\frac{-\nu\theta-\widetilde{\alpha}^k T}{\sqrt{2BT}}\biggr) - \exp\biggl(\frac{-\widetilde{\alpha}^k\nu\theta}{B}\biggr)\Phi\biggl(\frac{\nu\theta-\widetilde{\alpha}^k T}{\sqrt{2BT}}\biggr)
\end{equation}

Please note that in depleting environments, $P^k(t)$ and $Q^k(t)$ can also be computed through successive convolutions, but without the above simplification through the Laplace transform, since the drift term is different every time (see Appendix~\ref{App1}).\\

Fig.~\ref{fig:Panel2}(b)~(Left) shows an example of these analytical distributions $P^k(t)$, $Q^k(t)$, together with the distributions from the numerical simulations. Both $P^k(t)$ and $Q^k(t)$ display damped oscillations. Agents first leave their initial patch $0$ to go in the patch $1$, then go back, and so on. The widening of these distribution peaks is linked to the progressive desynchronization of agents. After an oscillatory period where the movements between the two patches are still coherent, stationary values, $P^k_\text{eq}$, $Q^k_\text{eq}$, are reached.\\

Below are metrics used to characterize the time spent in a patch (exploitation), the damping effect of these oscillations (cohesion) and their equilibrium values (accuracy) are presented. They will be used in the following sections.

\begin{itemize}
\item \textit{Exploitation} In this non-coupling case, the oscillations are periodic with period $\overline{T}^0+\overline{T}^1$, where $\overline{T}^k= -\frac{\theta}{\widetilde{\alpha}^k}$. Since the peaks are shifted between the two patches, we write that the ``mean'' departure time of each peak $\nu$ in a patch $k$ is 
\begin{equation}
    \overline{t}^k_\nu = \nu\overline{T}^0+(\nu-1+k)\overline{T}^1+(2\nu-2+k)T_\text{tr}
    \label{Tnuk}
\end{equation}
for $\nu=\{1,2,...\}$. These periodic times are indicated by vertical lines in Fig.~\ref{fig:Panel2}(b).\\
If $p^0=p^1$, $\overline{t}_\nu^k = (2\nu-1+k)(\overline{T}+T_\text{tr})-T_\text{tr}$,
with
$\overline{T}=\overline{T}^0=\overline{T}^1$.\\

Fig.~\ref{fig:Panel2}(d)(Bottom, Left) shows that $\overline{T}^k$ (and accordingly $\overline{t}_\nu^k$) is increasing with larger reward probabilities $p^k$. This means more exploitation in better patches, while patches with a low reward rate are associated with less exploration.

\item \textit{Cohesion} The dampening of $P^k$ oscillations can be characterized in the case of $p^0=p^1$, using equation~\eqref{Pequal} (see Appendix~\ref{App1}). Rewriting the amplitude $P^k(t)$ for each peak $\nu$ gives

\begin{equation}
\begin{split}
   P^k_{A,\nu}=\sum_{m=1}^\infty & \frac{2m-1+k}{\sqrt{4\pi(2\nu-1+k)^3}}
   \\& \times\frac{\widetilde{\alpha}^2}{\gamma B}\exp\biggl(-\frac{ (\nu-m)^2\gamma^2}{2\nu-1+k}\biggr)
\end{split}
\label{Pamp}
\end{equation}
Please note that for $\nu \xrightarrow{}\infty$,  
$P_{A,\nu}^k$ tends to the equilibrium value $P^k_\text{eq}$. This value is defined below.\\
$\gamma$, named here the damping parameter, is defined as

\begin{equation}
    \gamma = \sqrt{\frac{\widetilde{\alpha}}{B}\biggl(\widetilde{\alpha}T_\text{tr}-\theta\biggr)}
    \label{gamma}
\end{equation}

The smaller $\gamma$ values are, the more damped the oscillations are. For small differences between $p^0$ and $p^1$, $\gamma$ can be estimated using the average effective drift term $\frac{\widetilde{\alpha}^0+\widetilde{\alpha}^1}{2}$. 
\\

Fig.~\ref{fig:Panel2}(c) shows the amplitudes of each peak in equation~\eqref{Pamp} and the impact of the reward probability $p$ and travel time $T_\text{tr}$ values on the damping parameter $\gamma$. Larger reward probabilities $p$ lead to decreasing $\gamma$ values, which is correlated to more damped oscillations. Indeed unit distributions $\Psi^k_\nu$ have a larger variance for bigger $p$. Inversely, increasing travel times $T_\text{tr}$ lead to less damped oscillations, as the $\Psi^k_\nu$ are more spaced. \\

\item \textit{Accuracy} Finally, the distributions can be characterized by their equilibrium values $P_\text{eq}^k$, $Q_\text{eq}^k$.\\ The fraction of agents in a patch at equilibrium $Q_\text{eq}^k$ can be found after computing the entering and leaving fluxes (see Appendix~\ref{App2}).
\begin{equation}
    Q_\text{eq}^k = \frac{1}{-\frac{2\widetilde{\alpha}^k T_\text{tr}}{\theta} + 1 + \frac{\widetilde{\alpha}^k}{\widetilde{\alpha}^{k'}}}
    \label{Qeq}
\end{equation}
And $P_\text{eq}^k$ can be estimated as
\begin{equation}
    P_\text{eq}^k = \frac{Q_\text{eq}^k}{\overline{T^k}}=\frac{\widetilde{\alpha}^k}{2\widetilde{\alpha}^k T_\text{tr} - \theta\biggl(1 + \frac{\widetilde{\alpha}^k}{\widetilde{\alpha}^{k'}}\biggr)}
    \label{Peq}
\end{equation}
\end{itemize}

Fig.~\ref{fig:Panel2}(d) shows the impact of changing the reward probabilities $p^0$, $p^1$, and the travel time $T_\text{tr}$ on the equilibrium values in the patch $1$, $P_\text{eq}^1$, $Q_\text{eq}^1$, $\overline{T}^1$. This highlights that the bigger the ratio $p^1/p^0$ is, the more important the fraction of agents in the patch 1 is. And inversely for the patch 0. Moreover, the higher the reward probabilities are, the smaller the probability to leave the patch and the larger the mean patch residence time is. Finally, increasing traveling times $T_\text{tr}$  diminish $Q_\text{eq}^k$, meaning that a bigger fraction of agents are traveling and not in patches. It also leads to smaller $P_\text{eq}^k$, which can be also explained by the increased number of agents traveling.\\

\begin{figure*}
    \centering
    \includegraphics[width=16cm]{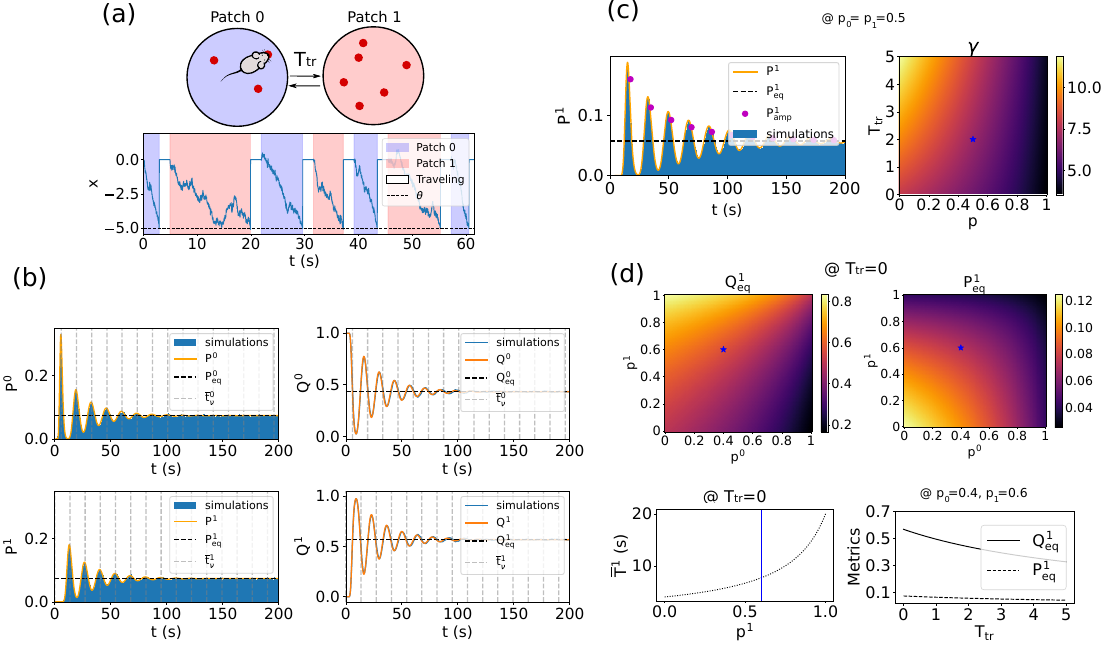}
    \caption{{\textbf{Characterization of the collective dynamics for non-interacting agents in two non-depleting patches}}.
    (a) Schema of the two patches environment. Foraging agents accumulate evidence in a patch of food (e.g. patch 0, shaded in blue), and leave it when their decision variable $x$ reaches the threshold $\theta$. They then go to the other patch (e.g. patch 1, shaded in red), after a travel time (non-shaded area). This process is repeated during the duration of the experiment.
    (b) Distributions of leaving times $P^k(t)$ (Left), and fraction of agents $Q^k(t)$ (Right) in patches $k=0$ (Top) and $k=1$ (Bottom) show oscillations converging to an equilibrium state ($2\times 10^4$ simulations). (c) Damped oscillations quantification. (Left) $P^k_{A,\nu}$ amplitudes diminish with time ($2\times 10^4$ simulations). (Right) The damping parameter $\gamma$ decreases with $p$ and increases with $T_\text{tr}$. (d) Equilibrium states and mean residence time quantification. (Top, Left) Equilibrium fraction in the best patch $Q_\text{eq}^1$ increases with $p^1$, and decreases with $p^0$. (Top, Right) Equilibrium leaving probability of the best patch $P_\text{eq}^1$ decreases with $p^0$ and $p^1$. In this non-interacting situation, $P_\text{eq}^1=P_\text{eq}^0$. (Bottom, Left) Mean best patch residence time $\overline{T}^1$ increases with $p^1$. (Bottom, Right) $Q_\text{eq}^1$, $P_\text{eq}^1$ decrease with the travel time $T_\text{tr}$.}
    \label{fig:Panel2}
\end{figure*}

The three metrics (exploitation, cohesion, accuracy) detailed above are useful for studying more complex situations, such as a group of interacting agents (see section~\ref{socinfo}) or individuals inferring which patch has the highest reward probability.

\subsubsection{Effect of best patch inference}
\label{IndLearning}
So far, no individual long-term learning of which patch would have the highest reward rate has been taken into account. The impact of this internal inference is studied below. The average $y$ value at the departure time $t_{y,\nu}^k$ can be estimated with $\overline{y}_\nu^k$ (See Appendix~\ref{App3}).

For $\nu\in \{1,2,...\}$,
\begin{equation}
    y_\nu^0 = y_{\nu-1}^{1}\exp\Biggl(\frac{p^0\theta\bigl(y_b+\frac{1}{2}\bigr)}{\tau_y\bigl(\alpha\bigl(y_b+y_{\nu-1}^1\bigr)-p^0\bigl(y_b+\frac{1}{2}\bigr)\bigr)}\Biggr)
\end{equation}
\begin{equation}
    y_\nu^1 = 1+\bigl(y_\nu^{0}-1\bigr)\exp\Biggl(\frac{p^1\theta\bigl(y_b+\frac{1}{2}\bigr)}{\tau_y\bigl(\alpha\bigl(y_b+1-y_\nu^0\bigr)-p^1\bigl(y_b+\frac{1}{2}\bigr)\bigr)}\Biggr)
    \label{ymean}
\end{equation}
With $y_0^1 = \frac{1}{2}$.\\
The average best patch inference is $\overline{y_\nu} = \frac{y^0_\nu+y^1_\nu}{2}$, plotted with $\overline{t_\nu} = \frac{t^0_{y,\nu}+t^1_{y,\nu}}{2} $

\begin{equation}
\begin{cases}
    t^0_{y,\nu} = t^{1}_{y,\nu-1} -\frac{\theta\bigl(y_b+\frac{1}{2}\bigr)}{\alpha\bigl(y_b+y^1_{\nu-1}\bigr)-p^0\bigl(y_b+\frac{1}{2}\bigr)}\\
    t^1_{y,\nu} = t^{0}_{y,\nu} -\frac{\theta\bigl(y_b+\frac{1}{2}\bigr)}{\alpha\bigl(y_b+y^0_{\nu}\bigr)-p^1\bigl(y_b+\frac{1}{2}\bigr)}
\end{cases}
\end{equation}

From the condition $\alpha^k\omega^k_{y_\text{max}}-p^k>0$, with $y_\text{max}=1$, two limits are derived for $y_b$.
\begin{equation}
   \frac{p^1}{2(\alpha-p^1)}>y_b> \frac{2\alpha-p^0}{2(\alpha-p^0)} 
   \label{yblim}
\end{equation}

It is possible to compute $P^k(t)$, $Q^k(t)$ with the updated $\widetilde{\alpha}^k\times \omega^{k_i}_\nu$,
and $Q^k_{eq}$, $P^k_{eq}$ for $\tau_y>>1$.\\

\begin{figure*}
    \centering
    \includegraphics[width=16cm]{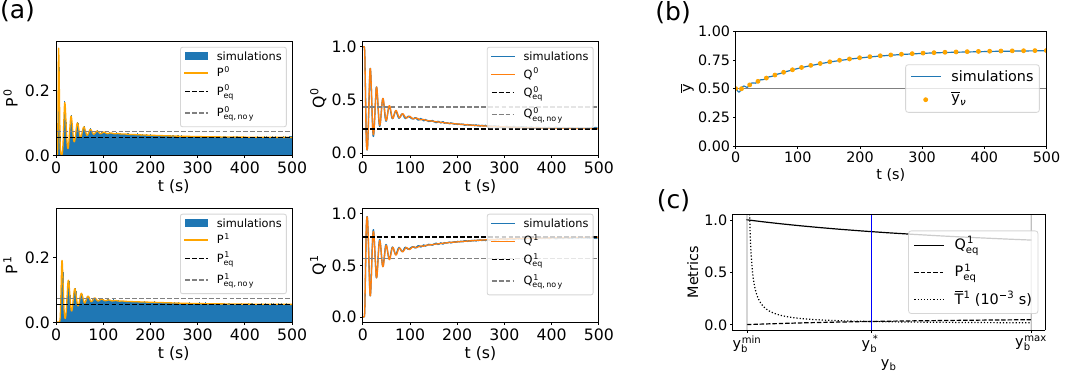}
    \caption{{\textbf{Best patch individual inference improves collective accuracy and exploitation times in the best patch}}. (a) Distributions of leaving times $P^k(t)$ (Left), and fraction of agents $Q^k(t)$ (Right) in patches $k=0$ (Top) and $k=1$ (Bottom) show damped oscillations and different equilibrium states compared to agents without best patch inference. $y_b=0.8$ ($2\times 10^4$ simulations). (b) The average best patch inference $\overline{y}(t)$ converges to a value $>\frac{1}{2}$, corresponding to a correct best patch inference ($k=1$). (c) Fraction in the best patch $Q_\text{eq}^1$ and mean patch residence time $\overline{T}^1$ decreases with $y_b$, while $P_\text{eq}^1$ increases. The two vertical gray lines represent the two limits in eq.~\eqref{yblim}. and the vertical blue line corresponds to the numerical $y_b$ value used in Fig.~\ref{fig:Panel22}(a). Please note that for all $y_b$ values, the equilibrium accuracy $Q_\text{eq}^1$ is higher than the situation without best patch inference (see Fig.~\ref{fig:Panel2}(b)).}
    \label{fig:Panel22}
\end{figure*}

Fig.~\ref{fig:Panel22} shows how distributions are changed with the best patch inference. The equilibrium fraction of agents in the best patch is progressively increasing as the mean best patch inference variable $\overline{y}$ is increasing. Fig.~\ref{fig:Panel22}(c) highlights a higher accuracy and patch residence times in the best patch, as well as a smaller $P_\text{eq}^1$ value for small boundary parameter $y_b$ values. Best patch individual inference improves collective accuracy and exploitation of the best patch.\\

The two previous sections were 
dedicated to non-interacting individuals, and helped understand the impact of non-social components on the collective dynamics. The next section will study how this dynamics is modified with social information sharing mechanisms between agents.

\subsubsection{Impact of social information}
\label{socinfo}
\begin{figure*}
    \centering
    \includegraphics[width=15cm]{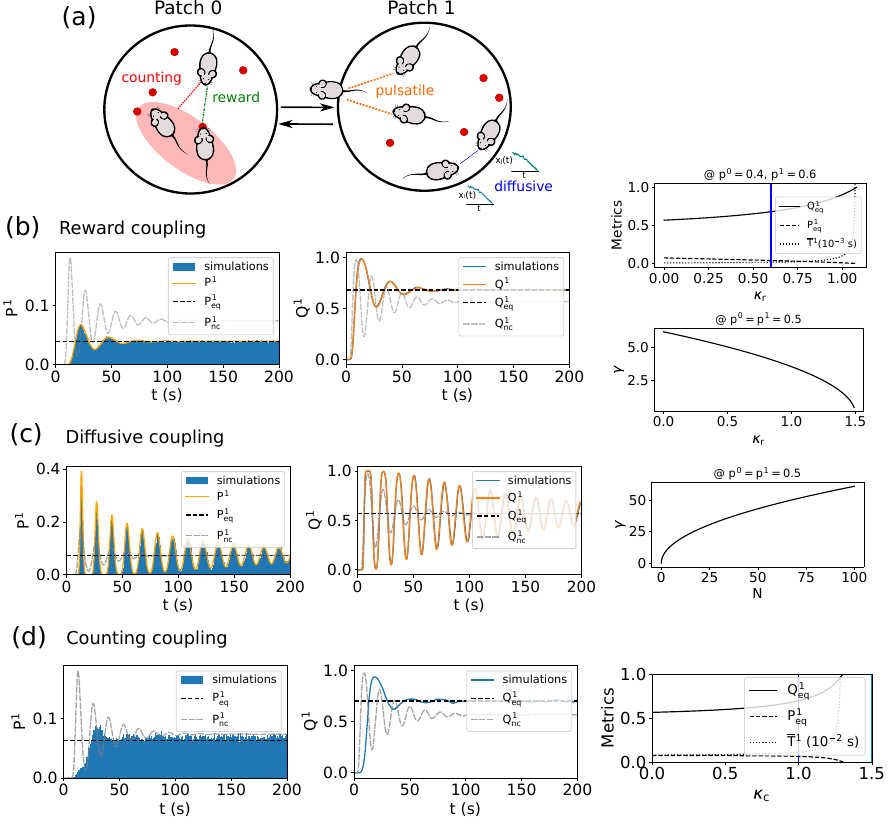}
    \caption{{\textbf{Collective social dynamics for reward, diffusive, and counting coupling in two non-depleting patches}}. (a) Schema of a group with interacting agents in a two-patches environment. Individuals interact with their conspecifics in different ways. Information can first come from the observation of others' food rewards (reward coupling). Individuals can also observe others' beliefs about a patch quality, by communicating continuously (diffusive), through agents arrivals or departures observation (pulsatile), or by observing the number of agents in their current patch (counting). (b) others' rewards observation increases accuracy, exploitation, and decreases cohesion. (Left) Distributions of leaving times $P^1(t)$, and fraction of agents $Q^1(t)$ in the best patch show more damped oscillations, later departure times, and an increased equilibrium accuracy. Parameters are $N=50$, $\kappa_r=0.6$ ($2\times 10^3$ simulations). (Right, Top) An increased $\kappa_r$ parameter leads to increased $Q_\text{eq}^1$, $\overline{T}^1$, and decreased $P_\text{eq}^1$ values. (Right, Bottom) The damping factor $\gamma$ decreases with $\kappa_r$. (c) Continuous belief sharing (diffusive) leads to more cohesion, with no accuracy and exploitation modifications. (Left) Distributions of leaving times $P^1(t)$, and fraction of agents $Q^1(t)$ in the best patch show less damped oscillations, and no change in equilibrium states and periodicity compared to a non-coupled situation. Parameters are $N=5$, $\kappa_\text{diff}=10$ ($2\times 10^3$ simulations). (Right) The damping parameter $\gamma$ increases with the group's size $N$. (d) Observation of the proportion of agents in a patch (counting) increases accuracy, exploitation, and reduces cohesion. (Left) Distributions of leaving times $P^1(t)$, and fraction of agents $Q^1(t)$ in the best patch show more damped oscillations, later departure times and a higher fraction of agents in the best patch. Parameters are $N=5$, $\kappa_\text{c}=1$ ($2\times 10^3$ simulations). (Right) Increased $\kappa_c$ values lead to higher $Q_\text{eq}^1$, $\overline{T}^1$, and decreased $P_\text{eq}^1$ values.}
    \label{fig:Panel3}
\end{figure*}

Fig.~\ref{fig:Panel3}(a) schematizes the different types of social information sharing. Their effects will be detailed below. For readability reasons, in this section the travel time is fixed to zero, $T_\text{tr}=0$ s.\\
A first information gathering mechanism can come from the observation of food
rewards that other agents in the same patch get.\\

\paragraph{Reward coupling}
To predict analytically the different distributions and metrics, it is possible to include the observation of others' food rewards (and their own rewards) in an effective drift term 
\begin{equation}
    \widetilde{\alpha}^k = \alpha - p^k(1+\kappa_r)
    \label{aeffkr}
\end{equation}

From the condition $\widetilde{\alpha}^k\geq 0$, the
upper limit for $\kappa_r$ is $\kappa_r \leq \text{min}\biggl(\frac{\alpha-p^0}{p^0},\frac{\alpha-p^1}{p^1}\biggr)$.\\

Fig.~\ref{fig:Panel3}(b) shows the impact of reward coupling on the analytical and simulated distributions $P^1(t)$, $Q^1(t)$, compared to the non-coupled ones. An increasing parameter $\kappa_r$ leads to a higher equilibrium fraction in the best patch $Q_\text{eq}^1$, a larger mean patch residence time $\overline{T}^1$, and a lower equilibrium leaving probability $P_\text{eq}^1$ linked to fewer departures from this patch 1. 
Moreover, this increasing reward coupling strength leads to a smaller damping parameter $\gamma$, i.e. to more damped oscillations. So, others' rewards observation increases accuracy, exploitation and decreases cohesion.\\

In addition to this public information about others' rewards, information about others' beliefs can also be collected. The different belief sharing processes are described below.\\

\paragraph{Diffusive coupling}
In the case of individuals sharing continuously their beliefs, it is possible to study the dynamics analytically in the strong limit of the coupling parameter $\kappa_\text{diff}$~\cite{bidariStochasticDynamicsSocial2022}. 

The mean $\overline{x} = \frac{\sum_{i=1}^N x_i}{N}$ and the half-difference between two agents $x^-_{ij} = \frac{x_i-x_j}{2}$
evolve as
\begin{equation}
    d\overline{x} = (p^k-\alpha)dt + \sqrt{2B}/N\sum_{i=1}^N dW_i
\end{equation}
\begin{equation}
    dx_{ij}^- = -\frac{\kappa_\text{diff}}{N}x_{ij}^-dt + \sqrt{B/2}(dW_i - dW_j)
\end{equation}
With the Ornstein-
Uhlenbeck process of $x_{ij}^-$, it is possible to show that for $\kappa_\text{diff}\xrightarrow{}\infty$, $\bigl<x_{ij}^-\bigr>=0$, $\bigl<(x_{ij}^-)^2\bigr>=0$~\cite{gardinerHandbookStochasticMethods1985}. Thus, in this limit all $x_i$ are considered equal, and the equivalent averaging equation can be described by
\begin{equation}
 d\overline{x} = (p^k-\alpha)dt + \sqrt{\frac{2B}{N}} dW(t)
 \label{xeqDiff}
\end{equation}
It is then possible to use the distributions described in the previous section with this new noise amplitude $B/N$. Please note that for $N>>1$, the noise term can be neglected and we only observe the mean residence times on the distributions. \\

Fig.~\ref{fig:Panel3}(c) shows the modification of the analytical and simulated distributions $P^1(t)$, $Q^1(t)$ with perfect diffusive coupling, compared to the non-coupled ones. As the noise amplitude $B$ has no effect on the equilibrium values $P^k_\text{eq}(t)$, $Q^k_\text{eq}(t)$ and $\overline{T}^k$ (see equations~\eqref{Peq}\eqref{Qeq}\eqref{T}), the effect of diffusive coupling can only be seen in the damping of oscillations.
An increasing number of agents $N$ leads to a higher damping parameter $\gamma$. Thus, the bigger a group is, the less damped the oscillations are. So, continuous belief sharing leads to more cohesion, with no accuracy and exploitation modifications. Normalization by agents in the observer's current or other patch highlights similar results, see supplementary Fig.~\ref{fig:Supdiff}.\\

Another way to infer others' beliefs about a patch's quality is to observe how many are in a patch.\\

\paragraph{Counting coupling}
For this coupling type, agents perceive the proportion of the group in their current patch. In this two patches environment, the mean counting parameter is $\eta = \frac{1}{2}$.\\ Analytical predictions of distributions are not possible in this coupling case, however, estimates of metrics at equilibrium can be calculated. For $N>>1$, the equilibrium effective equilibrium drift term can be written as

\begin{equation}
    \widetilde{\alpha}_\text{eq}^k = \alpha - p^k - \kappa_c\biggl(\frac{\widetilde{\alpha}_\text{eq}^{k'}}{\widetilde{\alpha}_\text{eq}^k+\widetilde{\alpha}_\text{eq}^{k'}}-\eta\biggr)
    \label{aeffSprop}
\end{equation}

Using the expression $Q_\text{eq}^k$ in eq.~\ref{Qeq}.\\
Solving the system of two equations ($k=0$, $k=1$) gives

\begin{equation}
\begin{split}
    \widetilde{\alpha}_\text{eq}^k = &\frac{2\alpha-(p^k+p^{k'})-\kappa_c(1-2\eta)}{2}\\&\times\biggl[1+\frac{(p^{k'}-p^k)}{2\alpha-(p^k+p^{k'})-2\kappa_c(1-\eta)} \biggr]
\end{split}
\end{equation}

The following limit can be derived from the condition $\widetilde{\alpha}_\text{eq}^k\geq 0$:
\begin{equation}
    \kappa_c\leq \frac{2\alpha - (p^k+p^{k'})}{2(1-\eta)}
    \label{limprop}
\end{equation}

Please note that without the condition $\widetilde{\alpha}_\text{eq}^k\geq 0$, in the limit $\kappa_c\xrightarrow{}\infty$, all agents stay in the best patch.\\

Fig.~\ref{fig:Panel3}(e)
shows the modification of the simulated distributions $P^1(t)$, $Q^1(t)$ with this counting coupling, compared to the non-coupled ones. It is not possible to compute the analytical distributions for counting coupling, and to have an expression for the damping parameter $\gamma$, but qualitative observations can still be made from simulated histograms. On these plots, oscillations are more damped with counting coupling, compared with the non-coupling case. Moreover, an increasing coupling parameter $\kappa_c$ leads to a higher fraction of agents in the best patch $Q_\text{eq}^1$, larger $\overline{T}^1$ and accordingly to a decreasing leaving probability $P_\text{eq}^1$. Observation of the proportion of agents in a patch increases accuracy, exploitation, and reduces cohesion. \\

Instead of looking at the number of neighbors in a patch, another way to collect information about others' beliefs is to observe their leaving and arriving movements.\\

\paragraph{Pulsatile coupling}
In this case agents communicate through pulses of departure or arrival information.\\

Analytical predictions are possible only for perfect pulsatile coupling ($\kappa_d\xrightarrow{}\infty$). In that case, all agents move together, the first one to leave is immediately followed by the others. So the number of agents in any patch $k$ is $n^k(t)=N$.
The unit distribution of patch residence times becomes
\begin{equation}
    \Psi_{\kappa_d\xrightarrow{}\infty}(T) = \Psi(T)\Omega(T)^{N-1}
\end{equation}
and can be used to compute $P^k(t)$, $Q^k(t)$ (see Appendix~\ref{App1}).\\

Otherwise, the effective drift term $\widetilde{\alpha}^k$ is time-dependent and predictable with difficulty~\cite{caginalpDecisionDynamicsGroups2017}\cite{karamchedBayesianEvidenceAccumulation2020}. However, equilibrium predictions are possible. The two normalization possibilities are studied below.\\

\begin{itemize}
    \item \textit{$\mathcal{N}_d=\mathcal{N}_a=N$:}\\
For a large number of agents $N>>1$, it is possible to write the equilibrium effective drift term as

\begin{equation}
    \widetilde{\alpha}_\text{eq}^k = \alpha - p^k + \frac{\kappa_a -\kappa_d}{\theta} \frac{\widetilde{\alpha}^k\widetilde{\alpha}^{k'}}{\widetilde{\alpha}^k+\widetilde{\alpha}^{k'}}
    \label{aeffSpulse2}
\end{equation}

Where $-\frac{\widetilde{\alpha}_\text{eq}^k}{\theta}$ are the stationary departure rates, and using Using the expression $Q_\text{eq}^k$ in eq.~\ref{Qeq}.\\ After solving the system of two equations, equation~\eqref{aeffSpulse} for $k=0$ and $k=1$, the equilibrium effective drift term can be written as

\begin{equation}
    \widetilde{\alpha}_\text{eq}^k = \frac{-a_2^k\pm \sqrt{(a_2^k)^2-4a_1a_3^k}}{2a_1}
    \label{aeffpulse2}
\end{equation}

With 
\begin{equation}
\begin{split}
    & a_1 = 2+\frac{\kappa_d-\kappa_a}{\theta}\\
    & a_2^k = \biggl(p^{k}-p^{k'}\biggr)\biggl(1+\frac{\kappa_d-\kappa_a}{\theta}\biggr) - 2\biggl(\alpha-p^k\biggr)\\
    & a_3^k=-\biggl(\alpha-p^k\biggr)\biggl(p^k-p^{k'}\biggr)
\end{split}
\end{equation}

Please not that $\widetilde{\alpha}_\text{eq}^k$ depends on the difference $\kappa_d-\kappa_a$ here.\\
From the condition $\widetilde{\alpha}_\text{eq}^k\geq 0$, 
the following relationship can be derived

\begin{equation}
    \kappa_a \leq \kappa_d+2\theta
    \label{limPulse2}
\end{equation}

The second normalization condition is treated alike.

\item \textit{$\mathcal{N}_d = n^{k_i}-1$, $\mathcal{N}_a = N-n^{k_i}-1$:}\\ 

Similarly, for a large number of agents $N>>1$ it is possible to write the equilibrium effective drift term as

\begin{equation}
    \widetilde{\alpha}_\text{eq}^k = \alpha - p^k - \kappa_d \frac{\widetilde{\alpha}_\text{eq}^k}{\theta}+ \kappa_a \frac{\widetilde{\alpha}_\text{eq}^{k'}}{\theta}
    \label{aeffSpulse}
\end{equation}

After solving the system of two equations, equation~\eqref{aeffSpulse2} for $k=0$ and $k=1$, the equilibrium effective drift term can be written as

\begin{equation}
    \widetilde{\alpha}_\text{eq}^k = \frac{1}{2}\biggl[\frac{2\alpha-(p^k+p^{k'})}{1+(\kappa_d-\kappa_a)/\theta}-\frac{(p^k-p^{k'})}{1+(\kappa_d+\kappa_a)/\theta}\biggr]
    \label{aeffpulse}
\end{equation}

From the condition $\widetilde{\alpha}_\text{eq}^k\geq 0$, 
the following relationship can be derived

\begin{equation}
    \kappa_a \leq -(\theta+\kappa_d)\times\text{min}\biggl(\frac{\alpha-p^{1}}{\alpha-p^0},\frac{\alpha-p^{0}}{\alpha-p^1}\biggr)
    \label{limPulse}
\end{equation}


And, since $\kappa_a\geq 0$, the upper limit for $\kappa_d$ is
\begin{equation}
    \kappa_d \leq |\theta|
    \label{limPulseb}
\end{equation}

Please note that in the case of $\kappa_d>0$ and $\kappa_a=0$, $Q^k_\text{eq}$ is independent of $\kappa_d$.
\end{itemize}

\begin{figure*}
    \centering
    \includegraphics[width=15cm]{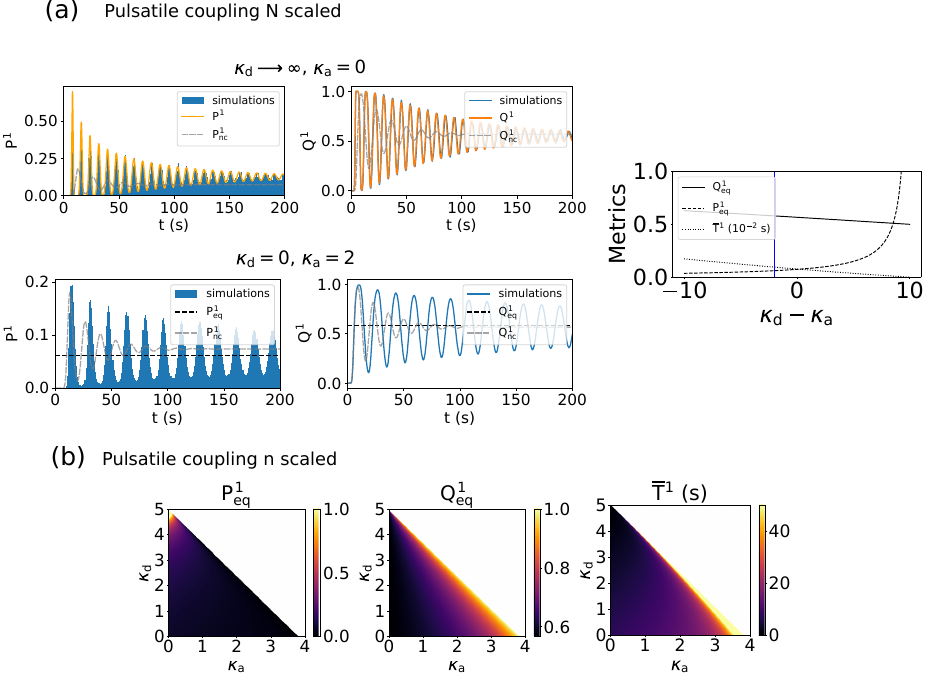}
    \caption{{\textbf{Collective social dynamics for pulsatile couplings in two non-depleting patches}}. (a) Pulsatile coupling scaled by the group's size $N$. (Left, Top)
    Observation of others' departures leads to less accuracy, less exploitation and more cohesion. Distributions of leaving times $P^1(t)$, and fraction of agents $Q^1(t)$ in the best patch show less damped oscillations, earlier departure times and a smaller equilibrium fraction of agents in the best patch. Parameters are $N=50$, $\kappa_\text{d}=100$, $\kappa_a=0$ ($1\times 10^3$ simulations). (Left, Bottom)
    Observation of others' arrivals leads to more accuracy, more exploitation, and more cohesion. Distributions of leaving times $P^1(t)$, and fraction of agents $Q^1(t)$ in the best patch show less damped oscillations, later departure times, and a higher equilibrium fraction of agents in the best patch. Parameters are $N=50$, $\kappa_\text{d}=0$, $\kappa_a=2$ ($1\times 10^3$ simulations).
    (Right) An increased $\kappa_d-\kappa_a$ difference leads to higher $P_\text{eq}^1$ and decreased $Q_\text{eq}^1$, $\overline{T}^1$ values. The blue vertical line represents the value used in Fig.~\ref{fig:Panel3_2}(a) Bottom. (b) Pulsatile coupling scaled by the number of other agents in the observer's current or other patch. An increased $\kappa_d$ parameter leads to higher $P_\text{eq}^1$, $Q_\text{eq}^1$ and decreased $\overline{T}^1$ values, except close to the upper boundary where the tendency is reversed for $P_\text{eq}^1$ and $\overline{T}^1$. For readability reasons $P_\text{eq}^1$ and $\overline{T}^1$ values are bounded respectively to $1$ and $50$s, and the upper part corresponds to sets of parameters beyond the limits~eq.~\eqref{limPulse}\eqref{limPulseb}.}
    \label{fig:Panel3_2}
\end{figure*}

Fig.~\ref{fig:Panel3_2}
shows the modification of the simulated distributions $P^1(t)$, $Q^1(t)$ with this pulsatile coupling, compared to the non-coupled ones. For a normalization by $N$, these plots highlight less damped oscillations. Higher $\kappa_d$ values compared to $\kappa_a$ are associated with a decreasing accuracy and mean patch residence time in the best patch, and a higher leaving probability. In that case, others' arrivals increase collective accuracy while departure perception is detrimental. For a normalization by the number of agents in a current patch, oscillatory behaviors are similar (see supplementary Fig.~\ref{fig:Supkdka}), but in that case, increased $\kappa_d$ values lead to an increased accuracy and patch residence time in the best patch. Observation of others' arrivals or departures lead to more cohesion. All in all, arrivals observation increases collective accuracy and exploitation, while departure observation diminishes exploitation, and may decrease or increase accuracy depending on the individual reference: the whole group, or only agents in the observer's current or other patch. \\

This two non-depleting patches environment represents a useful framework to study how different social information sharing mechanisms and social conditions lead to different effects on the group cohesion, exploitation, and global accuracy. In this environment the most advantageous strategy related to food intake is to spend the longest time in this best patch, to increase the total food intake. Another environment, successive depleting patches, introduces another aspect: agents have an additional interest to be in a patch with fewer individuals, to have access to more resources.

\subsection{Successive depleting patches}
\label{Succpatches}
In this section, agents go through $K$ successive patches of equal quality (same initial reward probabilities $p_\text{in}$), without going back to the previous patch (see Fig.~\ref{fig:Panel5}(a)).\\
Similar to the previous section, two different social conditions are studied successively in this section: non-interacting~\ref{nonIsucc} and interacting~\ref{Isucc} agents.
In particular, the first section below enables to define the distributions that are slightly different compared to the two patches section, to characterize the collective dynamics.

\subsubsection{Characterization of distributions for non-interacting agents}
\label{nonIsucc}

Depending on the non-depleting or depleting nature of patches, the analytical predictions are different.

\begin{itemize}
    \item For non-depleting patches, the total distribution of leaving times is
\begin{equation}
   P_\text{tot}(t) = \sum_{k=1}^K \Psi_{k}\bigl(t-(k-1) T_\text{tr}\bigr)
   \label{eq:PKnondep}
\end{equation}
and the fraction of agents in a patch $1<k<K$ is
\begin{equation}
\begin{split}
   Q^k(t) = \Omega_{k}\bigl(t-(k-1) T_\text{tr}\bigr)-\Omega_{k-1}\bigl(t-(k-1) T_\text{tr}\bigr)
\end{split}
\label{eq:QKnondep}
\end{equation}

With $Q^1(t) = \Omega_1(t)$, and\\ $Q^K(t)= 1-\Omega_{K-1}(t-(K-1)T_\text{tr})$ for the first and last patches.

\item For depleting patches with a single agent ($N=1$), the distributions can also be found, but without the above simplification. Indeed, in that case, the average reward rate is given by the time-dependent term in equation~\eqref{pdep}, which makes the simplification through the Laplace transform not possible. The probability distributions $P^k(t)$ are calculated through the successive convolutions, 

\begin{widetext}
\begin{equation}
\begin{split}
P^k(t) = & \int_0^{t-T_\text{tr}} d\tau_{k-1} \Psi^{k}(t-(\tau_{k-1}+T_\text{tr}))\int_0^{\tau_k-T_\text{tr}} d\tau_{k-2} \Psi^{k-1}(\tau_{k-1}-(\tau_{k-2}+T_\text{tr}))\\
... & \int_0^{\tau_3-T_\text{tr}}d\tau_2 \Psi^{3}(\tau_3-(\tau_2+T_\text{tr}))\int_0^{\tau_2-T_\text{tr}} d\tau_{1} \Psi^{2}(\tau_2-(\tau_1+T_\text{tr}))\Psi^{1}(\tau_1)
\end{split}
\label{Pappendix}
\end{equation}
\end{widetext}

and the total distribution of leaving times is $P_\text{tot}(t) = \sum_{k=1}^K P_{k}(t)$.

From these equations, the fraction of agents in a patch $k$ is, for $1<k<K$,
\begin{equation}
    Q^k(t) = \int_0^t d\tau [P^{k-1}(\tau-T_\text{tr})-P^k(\tau)] 
\end{equation}
With $Q^1(t) = 1-\int_0^t d\tau P^1(\tau)$, and\\ $Q^K(t)= \int_0^t d\tau P^{K-1}(\tau-T_\text{tr})$ for the first and last patches.\\
Please note that analytical predictions for $N>1$ in this non-interacting case are not possible, due to the complicated estimate of the average reward rate.
\end{itemize}

In the next section studying the modification of the collective dynamics within a social group, only depleting patches will be considered.

\subsubsection{Impact of social information}
\label{Isucc}
The effect of social information coupling in a group of foraging individuals is detailed here, with similar information sharing mechanisms described in the Methods section. Reward coupling is not detailed in this section, as it has been shown to correspond to an increased perceived reward probability.\\
First, the impact of actively shared signals about others' beliefs in this successive depleting patches environment is studied.\\

\paragraph{Diffusive coupling} 
Similar to the previous section, it is possible to study the dynamics with diffusive coupling analytically in the strong limit of the coupling parameter $\kappa_\text{diff}$. 
In that case, all agents move together, i.e. the number of agents in any patch $k$ is $n^k(t)=N$. The average reward rate is thus
\begin{equation}
\overline{p}(t) = p_\text{in}dt A_m \frac{1-e^{-\frac{N t}{A_m\Delta_r}}}{N t}
\label{pdepN}
\end{equation}

Here also, there is a decision variable equation equivalent to a non-coupled one, with the noise parameter scale by $N$: $B/N$.\\

\begin{figure*}
    \includegraphics[width=15cm]{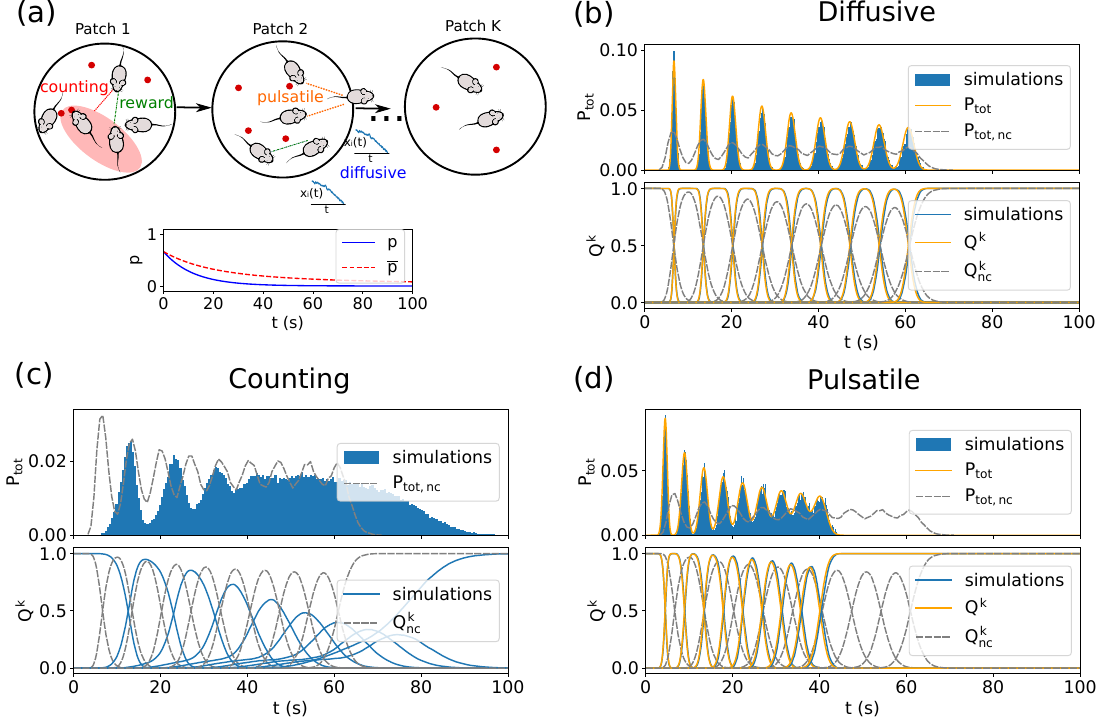}
    \caption{{\textbf{Collective social dynamics in successive depleting patches}}. (a) (Top) Schema of a  $K$ successive patches environment. Individuals all start from patch $1$, and move to the next patch without going back. The same information sharing mechanisms can be used to interact (reward, diffusive, pulsatile, counting). (Bottom) Reward probability $p(t)$ and average reward probability $\overline{p}(t)$ decrease exponentially in a perfectly cohesive group. (b) Continuous belief sharing (diffusive) leads to more cohesion, with no exploitation modifications. Distributions of leaving times $P_\text{tot}(t)$ (Top), and fraction of agents $Q^k(t)$ (Bottom) show less damped oscillations and no change in periodicity compared to a non-coupled situation. $\kappa_\text{diff}=100$ ($2\times 10^3$ simulations). (c) Observation of the proportion of agents in a patch (counting) increases exploitation and reduces cohesion. Distributions of leaving times $P_\text{tot}(t)$ (Top), and fraction of agents $Q^k(t)$ (Bottom) show more damped oscillations and later departure times. $\kappa_\text{c}=1$ ($2\times 10^3$ simulations). (d) Observation of others' departures leads to more cohesion and less exploitation. Distributions of leaving times $P_\text{tot}(t)$ (Top), and fraction of agents $Q^k(t)$ (Bottom) show less damped oscillations and earlier departure times. $N$ scaled, $\kappa_\text{d}=20$, $\kappa_a=0$ ($2\times 10^3$ simulations). }
    \label{fig:Panel5}
\end{figure*}

Fig.~\ref{fig:Panel5}(b) shows the modification of the analytical and simulated distributions $P_\text{tot}(t)$, $Q^k(t)$ with perfect diffusive coupling, compared to the non-coupled ones. An increasing number of agents $N$ leads to less damped oscillations, with no periodicity change. Similar to the previous section, continuous belief sharing leads to more cohesion, with no exploitation modifications. \\

Another way to infer others' beliefs about a patch's quality is to observe how many are in a patch.\\

\paragraph{Counting coupling}
Except for perfect counting case ($\kappa_c\xrightarrow{}\infty$), for which all agents stay in the initial patch, analytical distributions are not computable here. \\

Fig.~\ref{fig:Panel5}(c) 
shows the modification of the simulated distributions $P_\text{tot}(t)$, $Q^k(t)$ with counting coupling, compared to the non-coupled ones. These plots highlight more damped oscillations and later departure times. Similar to the previous section, observation of the proportion of agents in a patch increases exploitation and reduces cohesion.\\

Instead of looking at the number of neighbors in a patch, another way to collect information about others' beliefs is to observe their leaving and arriving movements.\\

\paragraph{Pulsatile coupling}
In this paragraph only the departure pulsatile coupling case is detailed, $\kappa_d>0$ and $\kappa_a=0$. Indeed as agents do not go back to a previous patch, the arrival information from such previous patch is less relevant here.

It is possible to study the dynamics with pulsatile coupling analytically in the strong limit of the coupling parameter $\kappa_\text{d}$. In that case, all agents move together, the first one to leave is immediately followed by the others. So the number of agents in any patch $k$ is $n^k(t)=N$. The average reward rate is thus the one written in equation~\eqref{pdepN}.
Similar to the two-patches section, the analytical distributions can be computed in this case.\\

Fig.~\ref{fig:Panel5}(d)
shows the modification of the simulated and analytical distributions $P_\text{tot}(t)$, $Q^k(t)$ with pulsatile coupling, compared to non-coupled ones. These plots highlight less damped oscillations, and earlier departure times. Observation of others' departures lead to more cohesion and reduce exploitation.\\

A successive depleting patches environment represents a useful framework to study how different information sharing mechanisms and social structures lead to different effects on the group cohesion and the exploration-exploitation trade-off.

\section{Discussion}
\subsection{Summary}
Group foraging is a collective behavior observed in a large variety of species~\cite{stephensForagingBehaviorEcology2007}. Its emergent dynamics have been studied through several quantitative models using various modeling approaches, such as the ideal free distribution~\cite{fretwellTerritorialBehaviorOther1969}, the  marginal value theorem for groups~\cite{livoreilPatchDepartureDecisions1997}, agent-based models~\cite{wajnbergOptimalPatchMovement2013}, reinforcement learning models~\cite{falcon-cortesCollectiveLearningIndividual2019}\cite{lofflerCollectiveForagingActive2023}, game-theoretic approaches~\cite{giraldeauSocialForagingTheory2000} \cite{cressmanGameTheoreticMethodsFunctional2014}, or bayesian models~\cite{perez-escuderoCollectiveAnimalBehavior2011}.
However, these models often do not take into account noise and environmental uncertainty, and are often
using general decision processes and interaction mechanisms, or can be too complex to derive analytical predictions. The model developed in this paper is based on a stochastic decision-making framework, evidence accumulation, that has already been applied to patch-leaving tasks~\cite{davidsonForagingEvidenceAccumulation2019}. The model presented here extends a two agent social foraging model ~\cite{bidariStochasticDynamicsSocial2022} by extendng it to larger groups to more forms of social information sharing mechanisms. The simple elements of the decision-making equation and the quasi-continuous food rewards keep the model mathematically tractable, which allows a better understanding of the links between individual properties and the emergent processes. Thus, through analytical analysis and numerical simulations, it has been possible to characterize how different social information sharing mechanisms lead to various collective dynamics.\\

The information sharing mechanisms implemented in this model represent different categories of social information~\cite{dallInformationItsUse2005}\cite{leadbeaterSocialLearningInsects2007}\cite{galefSocialInfluencesForaging2001}. For the two mechanisms previously described in~\cite{bidariStochasticDynamicsSocial2022}, diffusive coupling (continuous sharing of information) represents an actively shared signal, and pulsatile coupling (pulse of information at departure) represents a discrete social cue. In this paper, pulsatile coupling was further developed to include pulses of information at other agents' arrival. Another social cue mechanism was introduced, counting coupling (information about the number of other agents in the current patch). Finally, a public information mechanism was implemented, reward coupling (information about others' catches).\\

These social interaction mechanisms are studied within different environments, which allow studying different collective decision-making aspects. Three patch organizations are studied: a single patch, two non-depleting patches, and successive depleting patches. The first environment is useful to provide the basis of the different tools and metrics that are used in the two other cases. The metrics used to characterize a group's behavior are the probability of leaving a patch, and the fraction of agents in a patch. From these metrics, three main features are extracted, related to key social aspects: cohesion, accuracy, and exploitation~\cite{franksSpeedCohesionTradeoffs2013}\cite{stroeymeytImprovingDecisionSpeed2010}\cite{mehlhornUnpackingExplorationExploitation2015}\cite{monkHowEcologyShapes2018}.\\
A condition with all agents initially in the same patch enables the quantification of the group's cohesion. This can be observed with the damping of oscillations of the probability distributions, which can be quantified with a damping parameter $\gamma$ in some cases. The second aspect, accuracy, is here defined as the fraction of agents in the best patch. This can be evaluated in the two non-depleting patches environment with $Q^1_\text{eq}$.
Finally, in an environment with successive depleting patches, the time spent in a patch characterizes to what degree that patch is exploited. The mean residence time $\overline{T}$ is a useful metric to quantify this aspect.
Throughout this paper, numerical simulations and analytical analysis enabled characterizing how these different information sharing mechanisms and environments lead to various collective dynamics.\\

The different variations of non-social and social features in the model highlight diverse collective decision dynamics, that are summarized qualitatively in table~\ref{tabsum}.

\begin{table}[h!]
\begin{tabular}{|c|c|c|c|} 
 \hline
    Features & Cohesion $\gamma$ & Accuracy $Q^1$ & Exploitation $\overline{T}$\\ [0.5ex] 
 \hline
   Quality $p$ & $\searrow$ & $\nearrow$ (ratio) & $\nearrow$\\
  Depletion $A_m$ & $\nearrow$ & $\_$ & $\searrow$\\
  Travel $T_\text{tr}$ & $\nearrow$ & $\searrow$ (fraction) & $\O$\\
\hline
    Inference $y$ & $\_$ & $\nearrow$ & $\nearrow$\\
\hline
   Reward $\kappa_r$ & $\searrow$ & $\nearrow$ & $\nearrow$\\
  Diffusive $\kappa_\text{diff}$ & $\nearrow$ & $\O$  & $\O$ \\
 Pulsatile $\kappa_d$ & $\nearrow$ & $\searrow$ ($N$) or $\nearrow$ ($n^k$)  & $\searrow$\\
 Pulsatile $\kappa_a$ & $\nearrow$ & $\nearrow$ & $\nearrow$\\
 Counting $\kappa_c$ & $\searrow$ & $\nearrow$ & $\nearrow$\\
\hline
\end{tabular}
\caption{Classification of the effects ($\nearrow$ increasing, $\searrow$ decreasing, $\O$ no effect, $\_$ not relevant) of the environment (Patch quality, depletion, travel time), individual (best patch inference) and information sharing mechanisms (reward, diffusive, pulsatile, counting) on emergent collective characteristics (cohesion, accuracy, exploitation). This accuracy column is only relevant for the two non-depleting patches section.}
\label{tabsum}
\end{table}

\subsection{Effects of the environment and individual long-term learning on the collective dynamics features}
Our study highlights that environments of different statistics lead to different collective dynamics of agents foraging there.\\
First, we observed that increased reward probabilities are linked with a decrease of cohesion, as the variance of patch residence times increases. In better patches, the number of agents is higher~\cite{mFoodCompetitionForaging1988}. Our model showed that when the ratio between the food reward probabilities increases, accuracy increases as it becomes easier to distinguish the two patches. This is qualitatively similar to the ideal free distribution~\cite{fretwellTerritorialBehaviorOther1969}, but here the ratio between the fractions of agents in patches is not equal to the quality ratio and depends on other factors such as the cost of foraging. In addition, the mean residence time decreases for an increased reward probability, i.e. animals exploit more a patch if food is abundant. This is consistent with the predictions of the marginal value theorem~\cite{charnovOptimalForagingMarginal1976}.\\
Another environmental feature, depletion of resources, leads to an increased cohesion, with more exploration. This is consistent with the marginal value theorem~\cite{charnovOptimalForagingMarginal1976}. We may note that in our model animals always end up exploring other patches, indeed even if patches do not deplete, exploration might be adaptive to increase information and reduce boredom~\cite{cohenShouldStayShould2007}\cite{mehlhornUnpackingExplorationExploitation2015}.\\
Finally, increased travel times between patches lead to an increased cohesion and a smaller fraction in the best patch, as more agents are between patches, but do not change the exploration-exploitation behaviors. As longer traveling times are associated with longer residence times~\cite{kacelnikPsychologicalMechanismsMarginal1992}, this effect would have to be added in future works.\\

Our modeling framework also showed that individual processes such as a long-term memory effect of best patch inference has a significant positive impact on the collective efficiency. Indeed, individual inference of which patch is the best in the two-patches environment increases collective accuracy. This is consistent with empirical observations that individual learning increases foraging success ~\cite{lemanskiEffectIndividualLearning2021}\cite{eliassenQuantifyingAdaptiveValue2009}.\\

This paper also studied how different interaction rules in a social context would modify the collective foraging dynamics.

\subsection{Effects of the information sharing mechanisms on the collective dynamics features}
The cooperative information sharing mechanisms studied here have different effects on global dynamics.\\

In particular, for a two non-depleting patches environment, it is more advantageous for the group to spend more time in the patch with the highest reward rate. In that case, cooperative behaviors are crucial to increase the fraction of agents in the best patch. All information sharing mechanisms studied here increase accuracy, except the diffusive coupling which has no effect. These findings are consistent with empirical observations that information sharing increases foraging success~\cite{martinez-garciaOptimizingSearchResources2013}\cite{dallInformationItsUse2005}.\\
The impact of these interaction rules between foragers was more diverse when looking at group cohesion. Indeed, cohesion was increased with diffusive and pulsatile couplings, and decreased with reward and counting couplings. As greater cohesion can be important in the presence of threats, since it provides an increased collective vigilance and risk dilution~\cite{siegfriedFlockingAntipredatorStrategy1975}\cite{powellExperimentalAnalysisSocial1974}, one might hypothesizes that animals would communicate with active social signals (diffusive) and pay more attention to the leaving and arriving decisions of others (pulsatile) in the presence of a predator. Inversely in the absence of threat, agents can focus more on increasing the group's accuracy, through collecting information about the environment by observing others' catches (reward) or by looking at others' beliefs regarding the best patch (counting), which have both a strong positive impact on accuracy, increase exploitation but decrease the group's cohesion. Interestingly the pulsatile and counting couplings do not produce similar effects on the group's features, suggesting that the way agents infer others' beliefs have a strong impact on the collective dynamics and thus one mechanism might be preferred to the other depending on the context (for example predation).
Furthermore, pulsatile departure and arrival couplings do not produce the same effects. Arrivals observation increases collective accuracy and exploitation, while departure observation diminishes exploitation, and may decrease or increase accuracy depending on the individual reference: the whole group, or only agents in the observer's current or other patch. It will be interesting to infer such underlying mechanisms from an observed global movement dynamics~\cite{michelenaEffectsGroupSize2009}.\\
Finally, a successive depleting patches environment raises also the question of how to balance exploration and exploitation~\cite{franksSpeedCohesionTradeoffs2013}\cite{mehlhornUnpackingExplorationExploitation2015}\cite{monkHowEcologyShapes2018}. Our model showed that while some cooperative information sharing mechanisms lead to more exploitation (reward, arrival pulsatile and counting couplings), others favor exploration (departure pulsatile coupling).
More precisely for reward coupling for example, the use of public information can improve the estimation of patch quality and prevent the underutilization of resources~\cite{valoneGroupForagingPublic1989}
\\

All these different collective behaviors might be more or less advantageous depending on the context, and the same group can display various group dynamics depending on the context.
Famous examples are the fission-fusion dynamics in macaques~\cite{amiciSocialInhibitionBehavioural2018} or in dolphins~\cite{wursigDuskyDolphinsFlexibility2014}, or locusts that can be gregarious or social~\cite{petelskiSynergisticOlfactoryProcessing2024}. Depending on the habitat~\cite{wursigDuskyDolphinsFlexibility2014}, or on the group's composition (youngs or not for example)~\cite{soratoEffectsPredationRisk2012}, the collective behavior modes are different and contextually modulated.\\

All in all, our theoretical framework of social patch foraging highlighted a diversity of dynamics emerging from the variations of a mechanistic model. This core structure could later be further developed to account for more diverse foraginf situations.


\subsection{Limitations and future directions}
Among the limitations of the framework presented here, one is constituted by the fact that all the theoretical distributions were derived using the approximation of a constant drift term. This was calculated with a continuous reward rate, which is also the assumption of the marginal value theorem~\cite{charnovOptimalForagingMarginal1976}. Such computations are not possible for more time-spaced and noisy food intakes~\cite{realRISKFORAGINGSTOCHASTIC1986}. However, analytical predictions such as the mean residence times or equilibrium rates, as well as numerical investigations are still possible.
Another limitation of the environments implemented in our model is that fixed travel times were considered, while these can be distributed~\cite{wajnbergOptimalPatchTime2006}. And as longer traveling times are correlated with longer residence times~\cite{kacelnikPsychologicalMechanismsMarginal1992}, the model should be modified to study this effect. Other environmental limitations include the patch configurations. Here three simple patch configurations were used: single, two, successive patches. An environment with multiple patches, where a leaving agent could go in any of them, would be an important further development to relate to existing experimental~\cite{robertsForagingRadialMaze1989}\cite{franksSpeedCohesionTradeoffs2013} and field~\cite{strandburg-peshkinSharedDecisionmakingDrives2015} work with multiple patches.\\
In addition to these simplified environments, the individual processes such as the long-term best patch learning have been kept simple in this model. Indeed the best patch inference only took into account individual information. More complex information integration could be later developed to also grasp social interactions.
\\

In addition to these environmental and individual constraints, the social organizations studied in this paper could be extended in many ways. Some information sharing mechanisms have been studied in this paper, but more could be added. Another social cue example is chemical trails left by foraging bees on the food patches they already exploited. These marks may lead other bees to avoid these flowers~\cite{goulsonForagingBumblebeesAvoid1998}. Also, under certain circumstances, the use of public information can be suboptimal and misleading~\cite{giraldeauPotentialDisadvantagesUsing2002}, or
foraging agents (songbirds) only use social cues about the location of potential food sources, but not their profitability~\cite{hillemannInformationUseForaging2020}. Also, as social cohesion can be counterproductive in some cases~\cite{stutzCohesivenessReducesForaging2018}, the model should be further developed to explain the possible detrimental effects of cooperative patch foraging. Further developments also include a complexification of the social organizations, to make them closer to real foraging groups. This model studied an egalitarian group, however, more complex social structures exist~\cite{bakerForagingSuccessJunco1981}\cite{leeForagingDynamicsAre2018}. Including hierarchies in the model is a next step for the social foraging framework. Another important possibility would be to study competitive behaviors~\cite{rantaCompetitionCooperationSuccess1993}\cite{lagueEffectsFacilitationCompetition2012}, to link our theoretical framework with producer-scrounger models~\cite{vickeryProducersScroungersGroup1991}.\\

\section{Appendix}

\subsection{Derivation of $P^k(t)$ and $P^k_{A,\nu}$}
\label{App1}

The probability to leave a patch $k$ at time $t$, $P^k(t)$, is given by the convolution over the $\Psi^0(t)$ and $\Psi^{1}(t)$.
\begin{widetext}
\begin{equation}
\begin{split}
P_\nu^0(t) = & \int_0^{t-T_\text{tr}} d\tau_{\nu-1} \Psi^{0}(t-(\tau_{\nu-1}+T_\text{tr}))\int_0^{\tau_\nu-T_\text{tr}} d\tau_{\nu-2} \Psi^1(\tau_{\nu-1}-(\tau_{\nu-2}+T_\text{tr}))\\
... & \int_0^{\tau_2-T_\text{tr}}d\tau_1 \Psi^{0}(\tau_2-(\tau_1+T_\text{tr}))\int_0^{\tau_1-T_\text{tr}} d\tau_{0} \Psi^{1}(\tau_1-(\tau_0+T_\text{tr}))\Psi^{0}(\tau_0)
\end{split}
\label{Pappendix}
\end{equation}
\end{widetext}

This probability can be found after using the Laplace transform of $\Psi$ to compute the convolution of functions with the same drift term in one compartment
Through the Laplace transform :
\begin{equation}
\widetilde{\Psi^k}(s) = \exp \biggl(\frac{\theta}{B}\biggl(\widetilde{\alpha}^k-\sqrt{\widetilde{\alpha}^{k^2}+2sB}\biggr)\biggr) 
\end{equation}

The successive $\nu$ convolutions over distributions with the same effective drift term lead to the term $\biggl(\widetilde{\Psi^k}(s)\biggr)^\nu = \exp \biggl(\frac{\nu\theta}{B}\biggl(\widetilde{\alpha}^k-\sqrt{\widetilde{\alpha}^{k^2}+2sB}\biggr)\biggr)$. Going back to the real domain:

\begin{equation}
P_\nu^0(t) = \bigl(\Psi_\nu^{0}* \Psi_{\nu-1}^1\bigr) (t-2(\nu-1)T_\text{tr})
\end{equation}

With
\begin{equation}
    \Psi^k_{\nu}(t)=\frac{-\nu\theta}{\sqrt{4\pi B t^3}} \exp \biggl(\frac{-(\nu\theta+\widetilde{\alpha}^k t)^2}{4Bt} \biggr)
\end{equation}
for $\nu\geq 1$. $\Psi^k_0(t) = 1$ for all $t$.\\
The same reasoning for the patch 1 gives

\begin{equation}
P_\nu^1(t) = \bigl(\Psi_\nu^{0}* \Psi_{\nu}^1\bigr) (t-(2\nu-1)T_\text{tr})
\end{equation}

$P_\nu^k(t)$ can be written compactly as
\begin{equation}
P_\nu^k(t) = \bigl(\Psi_\nu^{0}* \Psi_{\nu-1+k}^1\bigr) (t-(2\nu-2+k)T_\text{tr})
\label{Pnuapp}
\end{equation}

$P^k(t)$ is then obtained by the sum over $\nu$ from 1 to $\infty$ of equations~\eqref{Pnuapp}. \\

The amplitude series $P^k_{A,\nu}$ in the case of $p^0=p^1$corresponds to $P^k$ computed at times $\overline{t}_\nu^k$.\\Since $\overline{t}_\nu^k = (2\nu-1+k)(\overline{T}+T_\text{tr})-T_\text{tr}$, it is not convenient to simplify the equation because of the last $-T_\text{tr}$. A $T_\text{tr}$ shift of the distributions can be used to define an alternative threshold $\theta'$, such that $\overline{T}+T_\text{tr} = -\frac{\theta'}{\widetilde{\alpha}}$. So $\theta' = \theta -\widetilde{\alpha}T_\text{tr}$, and the alternative peak times are instead ${\overline{t}_\nu^k}' = (2\nu-1+k)(\overline{T}+T_\text{tr})$.
The resulting expression~\eqref{Pamp} enables extracting the $\gamma$ parameter~\eqref{gamma}.

\subsection{Computing $Q_\text{eq}^k$ with entering and departing fluxes}
\label{App2}
We compute the balance of four agent fractions at equilibrium:
In patch 0 ($Q^0_\text{eq}$), in patch 1 ($Q^0_\text{eq}$), traveling from 0 to 1 ($W^{01}_\text{eq}$), and traveling from 1 to 0 ($W^{10}_\text{eq}$). These terms are linked by 
\begin{equation}
    Q^0_\text{eq}+Q^1_\text{eq}+W^{01}_\text{eq}+W^{10}_\text{eq}=1
    \label{frac}
\end{equation}
With the equilibrium departure rates equal to $-\alpha_\text{eff,eq}/\theta$, the balance system can be written as:

\begin{equation}
    \begin{cases}
        -\frac{\alpha^0_\text{eff,eq}}{\theta}Q^0_\text{eq} + \frac{1}{T_\text{tr}}W^{10}_\text{eq} = 0\\
         -\frac{1}{T_\text{tr}}W^{10}_\text{eq} -\frac{\alpha^1_\text{eff,eq}}{\theta}Q^1_\text{eq} = 0\\
         -\frac{\alpha^0_\text{eff,eq}}{\theta}Q^0_\text{eq} + \frac{1}{T_\text{tr}}W^{10}_\text{eq} = 0\\
         -\frac{1}{T_\text{tr}}W^{10}_\text{eq} -\frac{\alpha^1_\text{eff,eq}}{\theta}Q^1_\text{eq} = 0
    \end{cases}
    \label{fracsys}
\end{equation}

Solving the system of equation~\eqref{frac} and equations~\eqref{fracsys} gives the general equation  

\begin{equation}
    Q_\text{eq}^k = \frac{1}{-\frac{2\widetilde{\alpha}^k T_\text{tr}}{\theta} + 1 + \frac{\widetilde{\alpha}^k}{\widetilde{\alpha}^{k'}}}
    \label{Qeq}
\end{equation}

\subsection{Computing the recurrence relation with $\overline{y}^k_\nu$}
\label{App3}
It is possible to solve equation~\eqref{y} using solutions for $k=0$ and $k=1$ to estimate the mean $y$ value when leaving a patch $k$ for the $\nu$ time. 

\begin{equation}
\begin{cases}
    y_\nu^0 = y_\nu^{1}\exp\biggl(\frac{p^0\overline{T}^0_\nu}{\tau_y}\biggr)\\
    y_\nu^1 = 1+(y_\nu^{0}-1)\exp\biggl(\frac{p^1\overline{T}^1_\nu}{\tau_y}\biggr)
\end{cases}
\label{ybef}
\end{equation}

With $\overline{T}_\nu^k = -\frac{\theta}{\alpha\omega_\nu^k-p^k}$
and 
\begin{equation}
\begin{cases}
    \omega^{0}_\nu = \frac{y_b+\overline{y}_\nu^1}{y_b+\frac{1}{2}}\\
    \omega^{1}_\nu = \frac{y_b+1-\overline{y}_\nu^0}{y_b+\frac{1}{2}}
\end{cases}
\end{equation}
Since $\omega^k_\nu$ in patch $k$ is updated with the last $\overline{y}_m^{k'}$ from the other patch $k'$. Writing the detailed terms in equation~\eqref{ybef} gives the recurrence relation in equation~\eqref{ymean}.\\  


\bibliography{SocialForaging}

\newpage
\section*{Supplementary materials}
\renewcommand{\theHfigure}{S\arabic{figure}}
\renewcommand{\thefigure}{S\arabic{figure}}
\setcounter{figure}{0}
\begin{figure}[h!]
    \centering
    \includegraphics[width=0.95\linewidth]{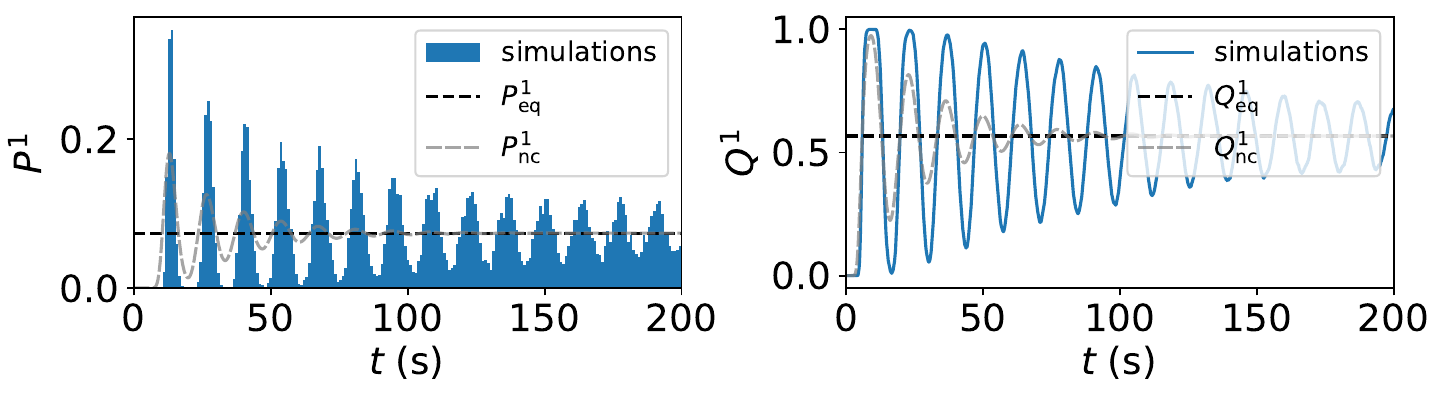}
    \caption{Diffuse coupling normalized by the number of other agents in the observer's current or other patch. Continuous belief sharing (diffusive) leads to more cohesion, with no accuracy and exploitation modifications. (Left) Distributions of leaving times $P^1(t)$, and fraction of agents $Q^1(t)$ in the best patch show less damped oscillations, and no change in equilibrium states and periodicity compared to a non-coupled situation. Parameters are $N=5$, $\kappa_\text{diff}=1$ ($2\times 10^3$ simulations).}
    \label{fig:Supdiff}
\end{figure}

\begin{figure}[h!]
    \centering
    \includegraphics[width=0.95\linewidth]{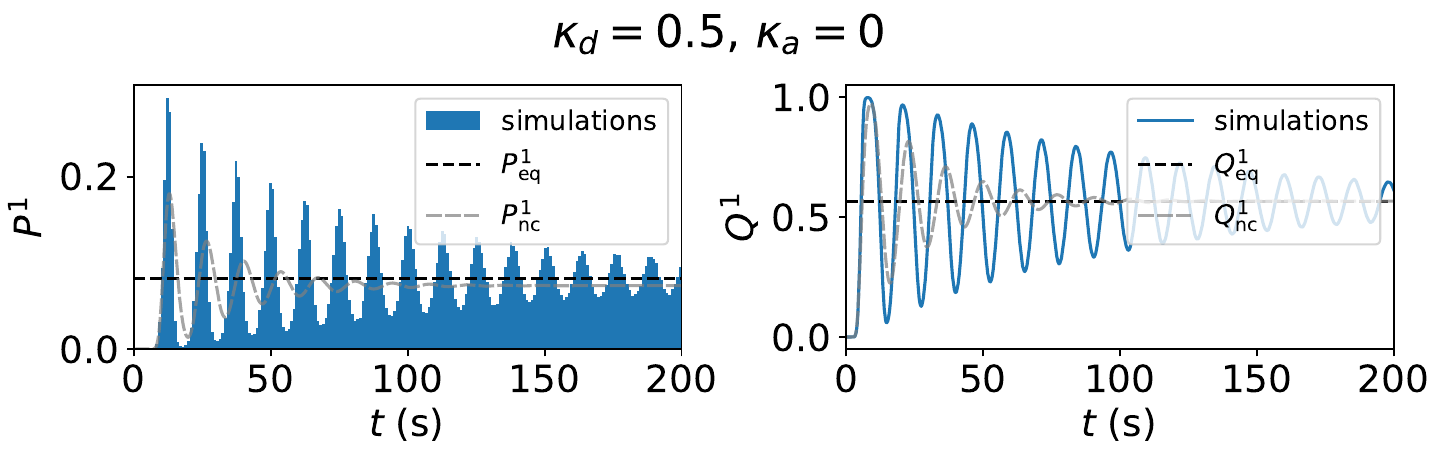}
    \includegraphics[width=0.95\linewidth]{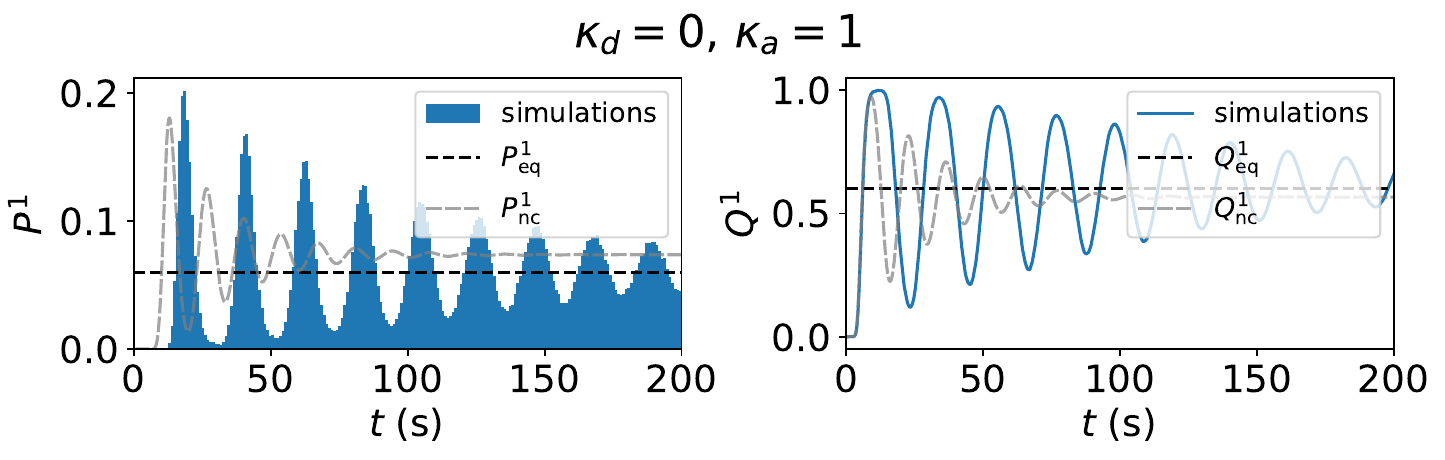}
    \caption{Pulsatile coupling scaled by the number of other agents in the observer's current or other patch. (Top) Observation of others' departures leads to more accuracy, less exploitation, and more cohesion. Distributions of leaving times $P^1(t)$, and fraction of agents $Q^1(t)$ in the best patch show less damped oscillations, earlier departure times, and a higher equilibrium fraction of agents in the best patch. Parameters are $N=50$, $\kappa_\text{d}=0.5$, $\kappa_a=0$ ($1\times 10^3$ simulations). (Bottom) Observation of others' arrivals leads to more accuracy, more exploitation, and more cohesion. Distributions of leaving times $P^1(t)$, and fraction of agents $Q^1(t)$ in the best patch show less damped oscillations, later departure times, and a higher equilibrium fraction of agents in the best patch. Parameters are $N=50$, $\kappa_\text{d}=0$, $\kappa_a=1$ ($1\times 10^3$ simulations).}
    \label{fig:Supkdka}
\end{figure}

\end{document}